\documentclass[10pt,journal,compsoc]{IEEEtran}
\IEEEoverridecommandlockouts

\usepackage[square,numbers,sort&compress]{natbib}

\usepackage{amsmath,amssymb,amsfonts}
\usepackage{graphicx}
\usepackage{textcomp}
\usepackage{xcolor}
\usepackage[colorinlistoftodos]{todonotes}
\usepackage{multirow}
\usepackage[tight,footnotesize]{subfigure}
\usepackage[linesnumbered,ruled]{algorithm2e}
\usepackage{algpseudocode}
\usepackage{enumerate}
\usepackage{xcolor}
\usepackage{slashbox}
\usepackage{makecell}
\usepackage{url}
\usepackage{breakurl}

\newcommand\mgape[1]{\gape{$\vcenter{\hbox{#1}}$}}

\def\BibTeX{{\rm B\kern-.05em{\sc i\kern-.025em b}\kern-.08em
    T\kern-.1667em\lower.7ex\hbox{E}\kern-.125emX}}
	
\graphicspath{{./figures/},{./skyplot/},{./photos/}}
\DeclareGraphicsExtensions{.pdf,.jpeg,.jpg,.png}
	
\begin{document}

\title{Adaptive SpMV/SpMSpV on GPUs for \\ Input Vectors of  Varied Sparsity}
\author{Min~Li,~\IEEEmembership{}
        Yulong~Ao,~\IEEEmembership{}
        and~Chao~Yang,~\IEEEmembership{Member,~IEEE}% <-this % stops a space
\IEEEcompsocitemizethanks{\IEEEcompsocthanksitem M. Li is with the Institute of Software,
Chinese Academy of Sciences, Beijing 100190, China, and University of
Chinese Academy of Sciences, Beijing 100049, China.
E-mail: limin2016@iscas.ac.cn.
\IEEEcompsocthanksitem Y. Ao is with the School of Mathematical Sciences, Peking University, Beijing 100871, China.
E-mail: aoyulong@outlook.com.
\IEEEcompsocthanksitem C. Yang (corresponding author) is with the School of Mathematical Sciences, Peking University, Beijing 100871, China,
and Peng Cheng Laboratory, Shenzhen 518052, China.
E-mail: chao\_yang@pku.edu.cn.}
\thanks{Nov 20, 2020. }}

\IEEEtitleabstractindextext{%
\begin{abstract}
Despite numerous efforts for optimizing the performance of Sparse Matrix and Vector Multiplication (SpMV) on modern hardware architectures, few works are done to its sparse counterpart, Sparse Matrix and Sparse Vector Multiplication (SpMSpV), not to mention dealing with input vectors of varied sparsity. The key challenge is that depending on the sparsity levels, distribution of data, and compute platform,
the optimal choice of SpMV/SpMSpV kernel can vary, and a static choice does not suffice.
In this paper, we propose an adaptive SpMV/SpMSpV framework, which can automatically select the appropriate SpMV/SpMSpV kernel on GPUs for any sparse matrix and vector at the runtime. Based on systematic analysis on key factors such as computing pattern, workload distribution and write-back strategy, eight candidate SpMV/SpMSpV kernels are encapsulated into the framework to achieve high performance in a seamless manner.
A comprehensive study on machine learning based kernel selector is performed to choose the kernel and adapt with the varieties of both the input and hardware from both accuracy and overhead perspectives. Experiments demonstrate that the adaptive framework can substantially outperform the previous state-of-the-art in real-world applications on NVIDIA Tesla K40m, P100 and V100 GPUs.

\end{abstract}

}

\maketitle

\IEEEdisplaynontitleabstractindextext

\IEEEpeerreviewmaketitle

\section{Introduction}
\label{intro}
The Sparse Matrix and Vector Multiplication (SpMV) is an essential building block
in a large variety of applications.
For example, in scientific computing, SpMV is a key primitive in solving large-scale sparse linear systems
that often arise from the discretizations of partial differential equations \cite{saad-book}.
In graph analytics, SpMV is frequently employed in many popular algorithms such as
breadth first search (BFS) \cite{bfs},
page rank \cite{pagerank}, triangles counting \cite{GraphMat-2015}, and
short-path finding \cite{GraphMat-2015}.
In machine learning, due in large part to the sparsity of the data samples and feature spaces,
SpMV is gaining increasingly more attention in the applications of, e.g.,
support-vector machine (SVM) and logistics regression.
Therefore, it is of crucial importance to study how to achieve high performance for SpMV
computations on modern hardware platforms.

Over the past few decades, extensive researches have been done to study the performance optimization of SpMV
on both CPUs \cite{cpu-spmv-1999-sc,cpu-spmv-2009,cpu-spmv-2010-cgo,cpu-spmv-2011-ipdps,pOSKI-2012,2018-CVR-spmv} and
GPUs \cite{Bell-2008-spmv, Ashari-sc14-spmv, Greathouse-sc14-spmv, CSR5, Merge-spmv,
naive-spmv, hola-spmv, new-format-CSX, new-format-clspmv, SMAT-spmv, dnn-spmv}.
With the advent of the big data era, the application scenarios are rapidly changing.
In traditional SpMV, the input vector is only assumed to be dense.
However, it has become increasingly common, especially in graph analytics and machine learning,
that the input vector is sparse and the sparsity may even vary during the program execution.
For instance, in the implementation of SVM {\cite{libsvm}}, the input vector is in fact generated from
a piece of data sample, i.e., a row of a sparse matrix, of which the number of nonzeros
may change because of the variation of matrix rows.
In BFS computing, the input vector often represents the current active vertices, which is also sparse
with varied sparsity from iteration to iteration.
In addition, similar computing patterns can also be found in other applications, such as the logistics regression
and the pruned fully-connected layer in deep learning networks.
It is a clear trend that the performance optimization of the sparse counterpart of SpMV,
the Sparse Matrix and Sparse Vector Multiplication (SpMSpV) is getting popularity.
So far, only very few efforts are seen for optimizing SpMSpV on CPUs
\cite{CombBLAS-2011,GraphMat-2015,GraphPad-2016,spmspv-bucket-2017}
and GPUs \cite{spmspv-sort-2015}.

In addition to the continuing efforts made to improve the performance of SpMV
and SpMSpV on GPUs, there are two major challenges arising from
the application and hardware development.
On the one hand,  the rapid change of GPU hardwares \cite{p100-whitepaper, v100-whitepaper}
makes it very hard, if possible at all, for a stand-alone SpMV/SpMSpV kernel
to be flexibly deployed on various GPUs.
On the other hand, the sparsity of the input vector can vary greatly in many applications,
and it is therefore very difficult to devise a single SpMV/SpMSpV kernel that works well
with all possible input situations. In this case, the adaptivity
between different sparse kernels is of great research and practical values.

However, it is not an easy task to design such an adaptive framework.
A major difficulty is how to identify suitable candidate kernels for the framework.
Although a few early studies \cite{direction-bfs, spmspv-push-pull-2018}
have brought up the idea of adaptivity by
considering the switch between a traditional SpMV kernel
and a high-performance SpMSpV kernel
as the candidates, deeper analysis is urgently required to answer the question:
are two kernels really enough?
Another important issue to be addressed is how to select an effective kernel
according to the characteristics of the input data.
The kernel selector should neither be too simple to lose accuracy and flexibility,
nor should it be too sophisticated to introduce large overhead.
A trade-off has to be made so as to design a kernel selector that is accurate, efficient
and easy to generalize.

In order to tackle these challenges, we propose a new adaptive framework that can automatically select the appropriate SpMV/SpMSpV kernel on GPUs for any sparse matrix and vector at the runtime.
To build the kernel  list, we carry out a series of analysis on key factors such as the computing pattern, workload distribution method and write-back strategy and identify eight candidate kernels.
Three machine learning models, instead of only one, are utilized to compose an accurate and lightweight
kernel selector that can deal with the complicated feature space arising from various application scenarios and hardware architectures.
The models are trained with 2,010 sparse matrices and a wide variety of input vectors
and then used to predict the most appropriate kernel on-the-fly.
Experiments show that the proposed adaptive SpMV/SpMSpV framework can work
well on three typical GPU platforms and substantially outperform the previous state-of-the-art
in both BFS and PageRank applications.
To the best of our knowledge, this is the first work to systematically
study the SpMV/SpMSpV kernel variants from both computing patterns and low-level
optimization technique perspecives and develop an adaptive framework for them on GPUs.

The remainder of the paper is organized as follows.
In Section~\ref{background} we discuss the background and motivation of this work.
Then an overview of the proposed adaptive SpMV/SpMSpV framework is presented
in Section~\ref{overview}.
Following that, details of the framework regarding the
candidate list, kernel selector and runtime procedure are respectively provided
in Section~\ref{candidate-kernel-exploration}, \ref{model} and \ref{runtime-process}.
Experiment results are presented and analyzed in Section~\ref{experiment}.
We briefly mention some related works in Section~\ref{related}.
And the paper is concluded in Section~\ref{conclusion}.

\section{Background and Motivation}
\label{background}

Both SpMV and SpMSpV compute the multiplication {between} a sparse matrix $A$ and a vector $x$,
where the matrix $A$ is of size $m \times n$ with $nnz$ nonzeros
and the vector $x$ is of length $n$.
The difference is that the former treats the vector $x$ as a dense input
while the latter deals with a sparse input vector $x$ with $nnz\_x$ nonzeros.
Considering the computing pattern, the existing SpMV/SpMSpV algorithms can be classified
into two categories: \emph{vector-driven} and \emph{matrix-driven} \cite{spmspv-bucket-2017}.
For vector-driven algorithms, the computation orders are decided by the values in the input vector.
On the other hand, the nonzeros of the sparse matrix determines the computation order
for matrix-driven algorithms.
We would like to point out that there is another way to classify existing SpMV/SpMSpV algorithms
by considering how the matrix is accessed, i.e., whether the nonzeros are accessed
along the row or column. Correspondingly, we call the SpMV/SpMSpV
algorithms \emph{row-major} and \emph{column-major} algorithms.

The access direction to the matrix entries may serve as a decisive factor
on how to select the storage format of the sparse matrix.
For the row-major approach, the matrix format should support random access
to any row of the matrix in $O(1)$ time, which can be naturally done by
using the broadly applied Compressed Sparse Row (CSR) format.
While for the column-major one, it is desirable to achieve fast random access to any column,
thus the Compressed Sparse Column (CSC) format is a suitable choice.
We remark that other sparse matrix storage formats could also be feasible in practice
and can be analogously incorporated into the proposed adaptive framework as needed.
The storage formats for the vector are straightforward:
one can store all elements of the dense vector in a single array for SpMV
or store only  the indices and values of the nonzeros of the sparse vector in two arrays for SpMSpV.
For sparse vectors, only a subset of nonzero entries of the sparse matrix can make a contribution to the final results.
For the convenience of expression, these nonzero entries are called the effective nonzeros of the matrix.
One can choose to either access only the effective nonzeros of the matrix to reduce the memory footprint,
or access all of them to avoid potential overhead due to value validation.

\begin{figure}[!htb]
\centering
\subfigure[BFS]{\includegraphics[width=0.48\columnwidth]{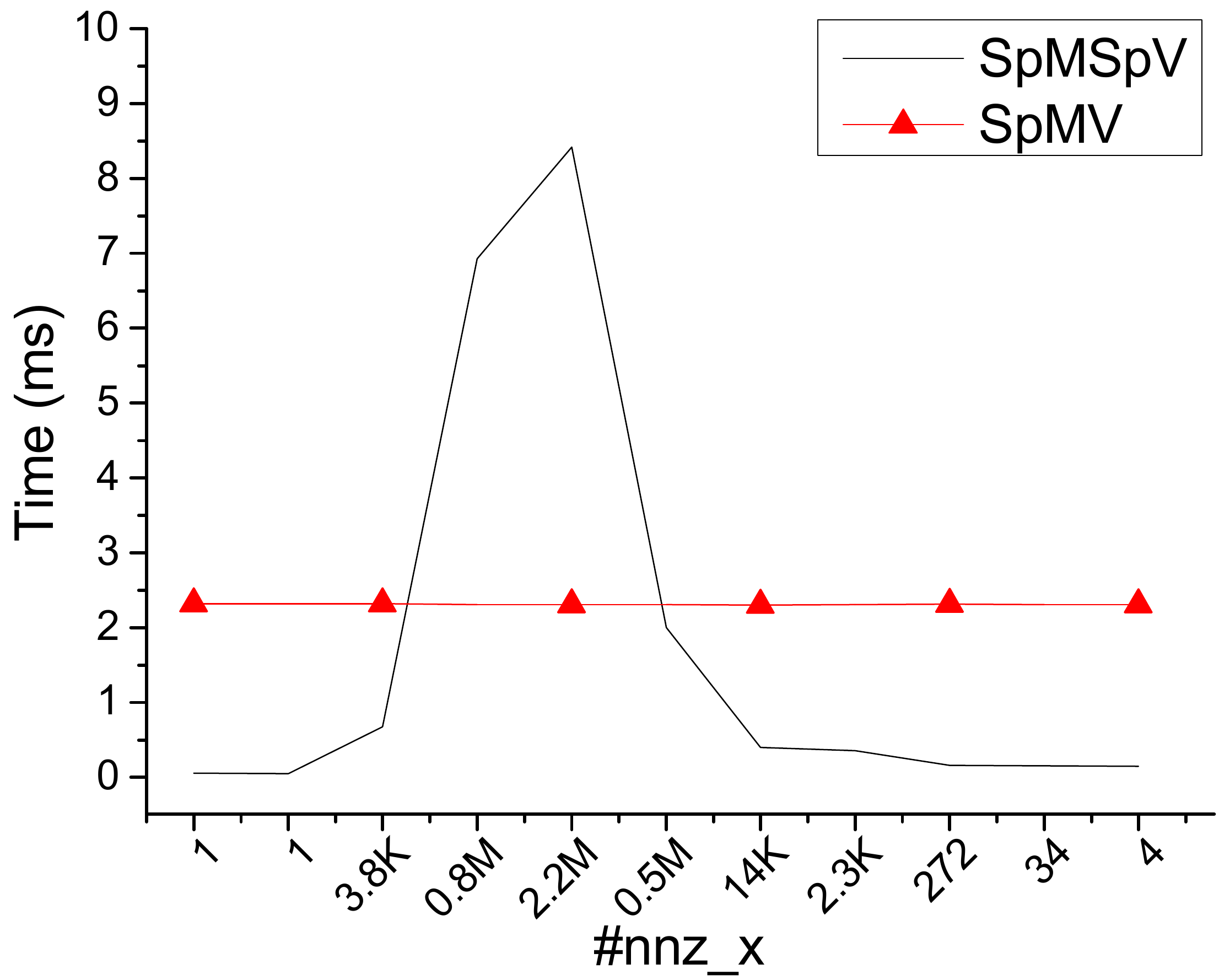}}
\subfigure[PageRank]{\includegraphics[width=0.48\columnwidth]{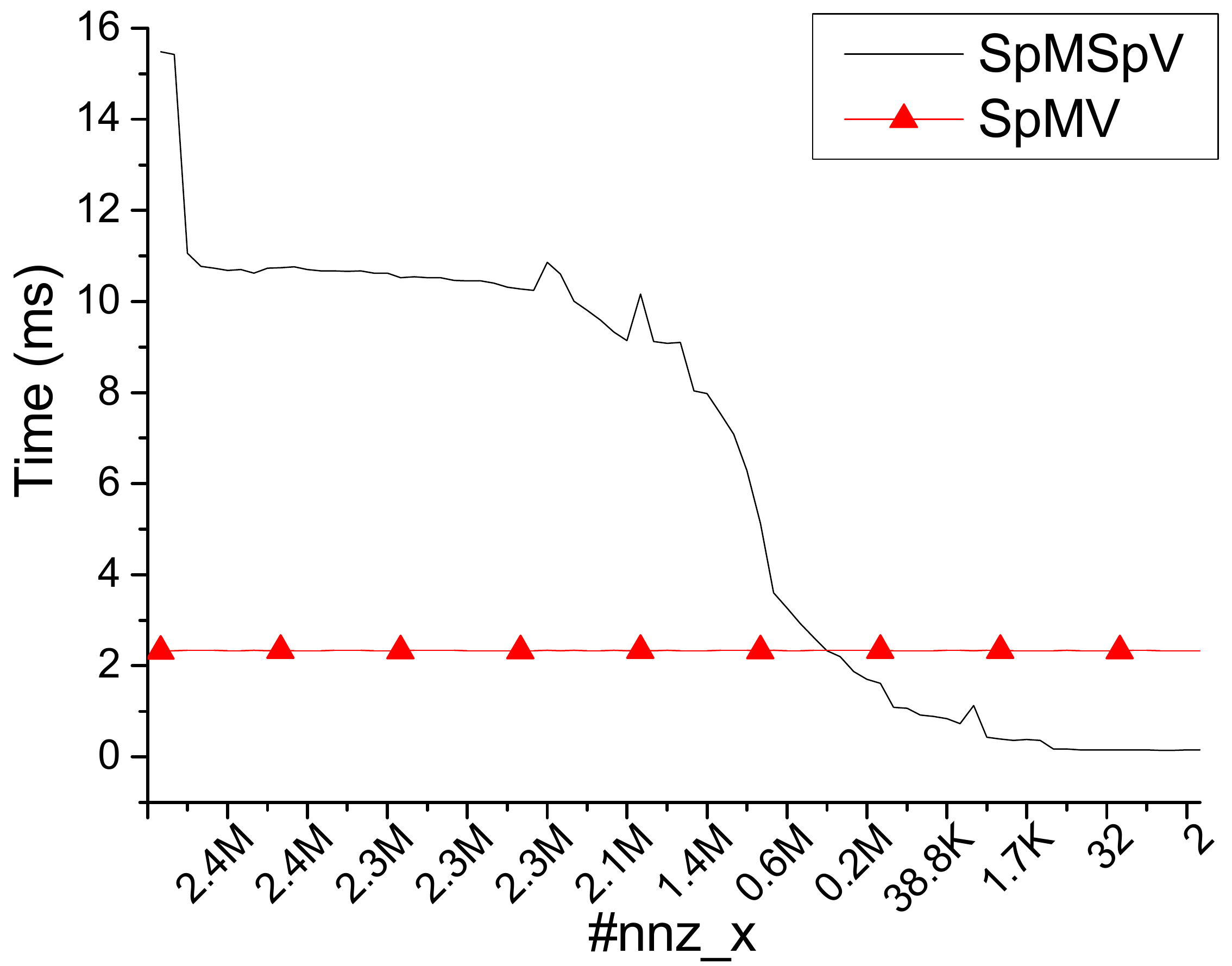}}
\caption{The performance of the SpMV \cite{hola-spmv} and SpMSpV \cite{spmspv-sort-2015} on the \textit{wikipedia-2007}
data set with the inputs vectors generated in real-world applications: (a) BFS, (b) PageRank. The 'K' and 'M' on the x-axis represent thousands and millions, respectively.}
\label{fig-motivation}
\end{figure}

For SpMV/SpMSpV, it is very common that the sparsity of the input vectors
varies during the execution of a real-world application.
In this case, a single SpMV or SpMSpV kernel does not suffice.
To show this, we provide an example with the inputs vectors generated by the BFS and
the incremental PageRank applications on the \textit{wikipedia-2007} data set.
The SpMV from \cite{hola-spmv} and the SpMSpV from \cite{spmspv-sort-2015}
are used as the target kernels. The test results are shown in Figure~\ref{fig-motivation}.
We can see from the figure that neither kernel is superior to the other for all cases.
If no adaptivity is introduced, there would be 1.3x-3.6x performance loss.
In addition to that, with the development of GPU architectures, advanced hardware features
are continuously introduced or enhanced,
such as hardware supported atomic instructions for floating-point numbers \cite{p100-whitepaper},
and independent thread scheduling \cite{v100-whitepaper}.
{These rapid changes} of the hardware features {make} it hard
for a high-performance kernel to be flexibly deployed on different GPUs.
There is an urgent need for an adaptive SpMV/SpMSpV framework
that can encapsulate many existing SpMV/SpMSpV kernels
and can efficiently adapt with both application and hardware varieties.

\begin{figure*}[!htb]
\centering
\includegraphics[width=1.98\columnwidth]{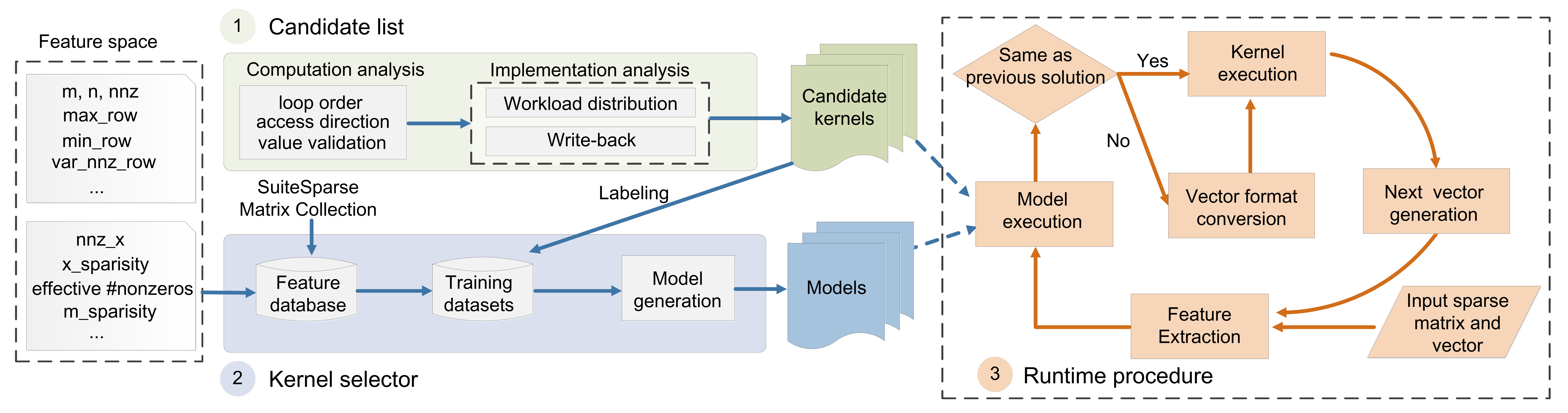}
\caption{The main components of the adaptive framework for SpMV/SpMSpV computations on GPUs.  }
\label{fig-framework}
\end{figure*}

\section{The Adaptive Framework}
\label{overview}

In this paper, we present an adaptive framework for SpMV/SpMSpV computations on GPUs.
The framework is illustrated in Figure~\ref{fig-framework},
which involves three major components:
\emph{candidate list}, \emph{kernel selector}, and \emph{runtime procedure}.

The candidate list is established and maintained as a list of candidate SpMV/SpMSpV kernels
that will be used by the adaptive framework.
These candidate kernels are chosen based on analysis of the computing pattern
and the implementation factors, and can be flexibly extended as needed.
In this work, from the computation perspective, we search the solution space
by considering factors including {loop order}, {access direction},
and {value validation}, leading to three potential SpMV/SpMSpV solutions.
Based on it, we further explore from the implementation aspect and analyze
how workload-distribution and write-back strategies would influence the performance.
As a result, a total of eight candidate kernels are chosen and put into the candidate list.
All of these candidate kernels can play a role under certain conditions,
and will serve as a basis of the adaptive framework.

The kernel selector is responsible for selecting the proper kernel from the candidate list.
There are several ways to design the kernel selector, such as using a heuristic model, a performance model, or a machine learning model.
To be more automated, easily maintained in the long run, and flexibly extended to other hardware platforms, we choose to use machine learning models in the current study.
To deal with various application scenarios and hardware architectures, we employ three machine learning models to select
the computing pattern, the workload distribution method and the write-back strategy, respectively,
and devise feature spaces that contain necessary characteristics of the input data.
A number of 2,010 sparse matrices from SuiteSparse matrix collection and a wide variety of input vectors
are utilized
to help constitute the training and {testing data sets}, with data labeled with the most suitable kernels.
By feeding the training data into the machine learning models, the rule-based models are generated
and then integrated into the adaptive framework as a key component.

The runtime procedure of the adaptive framework executes as follows.
At the beginning, the sparse matrix and the initial input vector are supplied to the framework.
As the application proceeds, the features of the input vectors are extracted on-the-fly,
and the pre-trained machine learning models are employed
to predict the most appropriate kernel.
If the newly predicted kernel is different from the previously selected kernel,
the adaptive framework will switch to the new one.
In this case data format conversion is triggered only when necessary.
The whole procedure repeats until finishing processing all input vectors.

In the following, we will present the technical details of the three main components
of the proposed adaptive SpMV/SpMSpV framework.

\section{Candidate List}
\label{candidate-kernel-exploration}

To establish the kernel candidate list, we consider key factors related to both computation and implementation.
The detailed analysis is presented as follows.

\subsection{Computation Analysis}
\label{solution-space}

\begin{figure}[!htb]
\centering
\includegraphics[width=0.92\columnwidth]{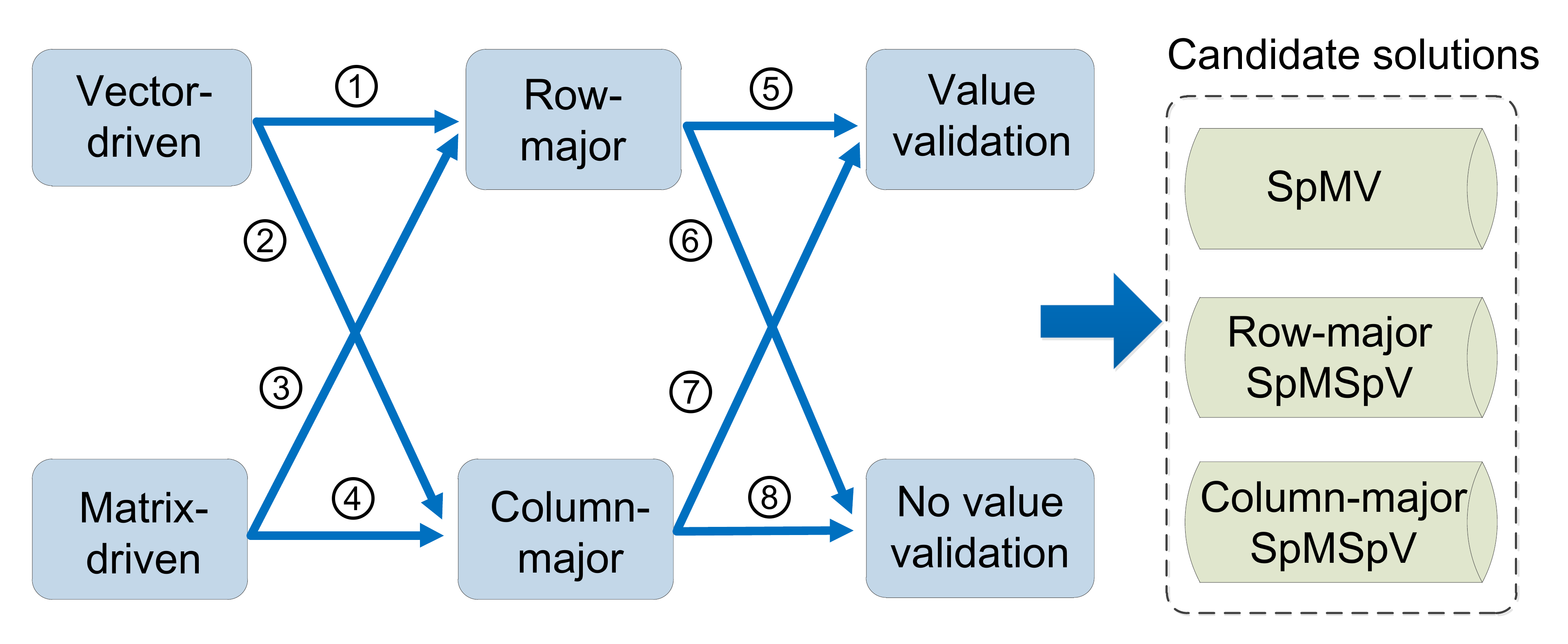}
\caption{Exploring the solution space from the computing perspective.  }
\label{fig-solution-space}
\end{figure}

The computation processes of SpMV and SpMSpV involve many choices,
such as the loop order, i.e., matrix-driven or vector driven, and the matrix access direction,
i.e., row-major or column-major.
In addition, there are also two choices to decide whether it is necessary or not
to validate if an entry of the matrix makes a contribution to the final results.
By combining the three aspects, there are in total eight possible solutions for SpMV and SpMSpV,
as depicted in Figure~\ref{fig-solution-space}.
In the figure, we tag different connections
among the available options with different numbers so that a possible solution can be represented by
a pair of numbers. For instance, $(\textcircled{1}, \textcircled{5})$ denotes the solution
that uses the vector-driven method, accesses the matrix values in a row-major manner,
and validates the effectiveness of the accessed matrix values.

We make further analysis on the above possible solutions and consider how to reduce the solution space.
When using a vector-driven approach,
if the matrix values are accessed in the row-major order, there are a large number of validations
required to check whether the computation needs the values.
On the other hand, if we access the matrix values in the column-major order,
the effective matrix entries can be extracted directly by the vector values without any validations.
Therefore, by contrast,  solution $(\textcircled{2}, \textcircled{8})$
is more efficient than {solutions $(\textcircled{1}, \textcircled{5})$,
$(\textcircled{1}, \textcircled{6})$ and $(\textcircled{2}, \textcircled{7})$}
for the vector-driven approach.
When it comes to a matrix-driven method, value validations are still needed
if matrix values are accessed in the column-major order.
This kind of computation is analogous to
$(\textcircled{2}, \textcircled{8})$, except for some extra {value validation} operations.
Therefore, matrix-driven column-major methods such as $(\textcircled{4}, \textcircled{7})$
and $(\textcircled{4}, \textcircled{8})$ are also ruled out.
When we access matrix values in the row-major order, we can decide whether value validation is needed or not.
The two resultant solutions, i.e.,  $(\textcircled{3}, \textcircled{5})$,
and $(\textcircled{3}, \textcircled{6})$, are both feasible choices.
In particular, it is worth noting that the matrix-driven column-major approach
may serve as a competitive method for multi-node environments of CPUs \cite{GraphMat-2015},
but is usually not suitable for GPUs due to the overhead introduced by extra value validations.
To sum up, we have three candidate solutions $(\textcircled{2}, \textcircled{8})$,
$(\textcircled{3}, \textcircled{5})$, and $(\textcircled{3}, \textcircled{6})$, that show potential
to achieve high performance. For the convenience of expression, $(\textcircled{2}, \textcircled{8})$ and
$(\textcircled{3}, \textcircled{5})$ are referred to
as column-major SpMSpV and row-major SpMSpV, respectively.
And $(\textcircled{3}, \textcircled{6})$ is in fact the traditional SpMV.

\begin{figure}[!htb]
\centering
\subfigure[\textit{ljournal-2008}]{\includegraphics[width=0.48\columnwidth]{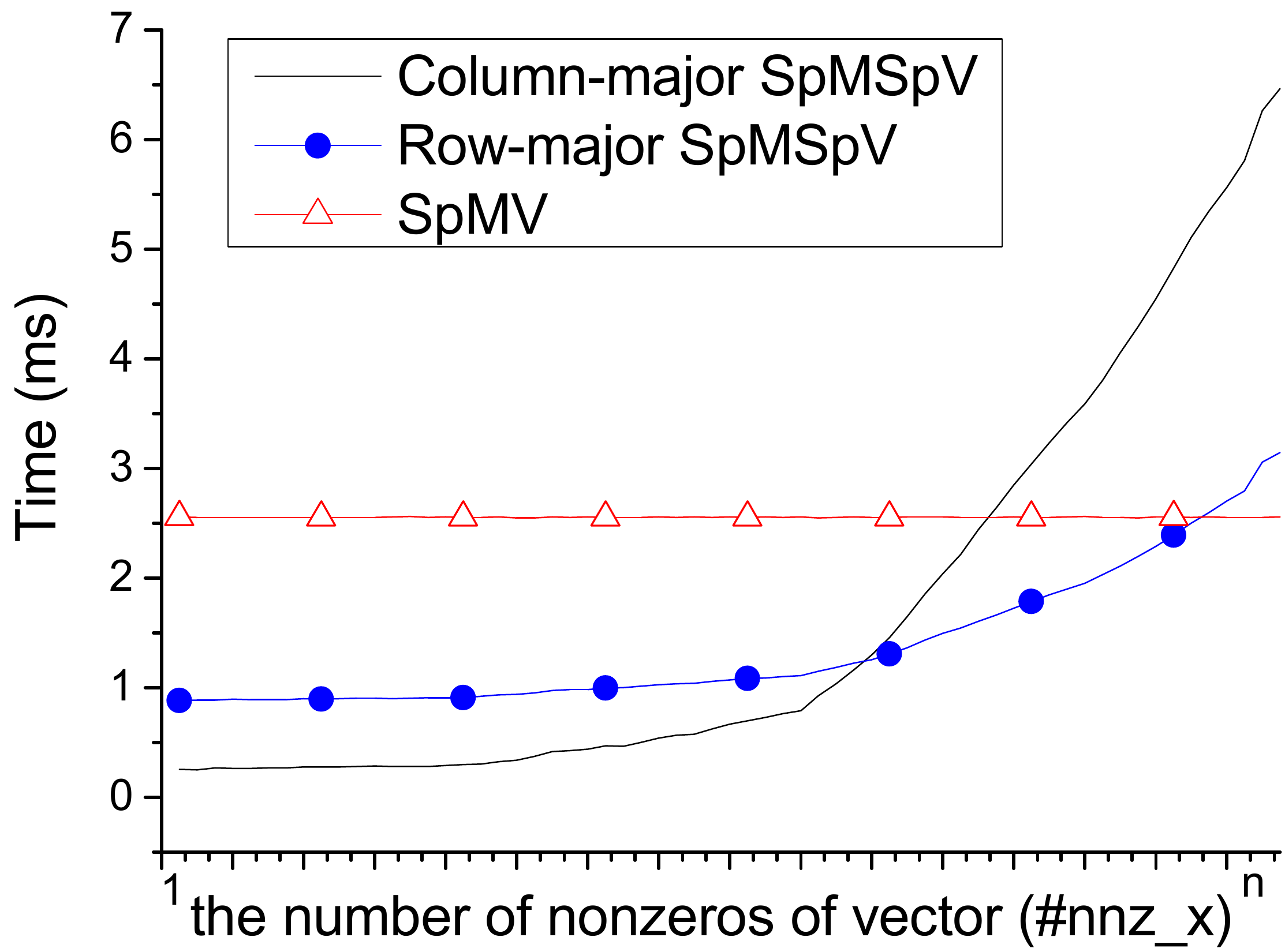}}
\subfigure[\textit{wikipedia-2007}]{\includegraphics[width=0.48\columnwidth]{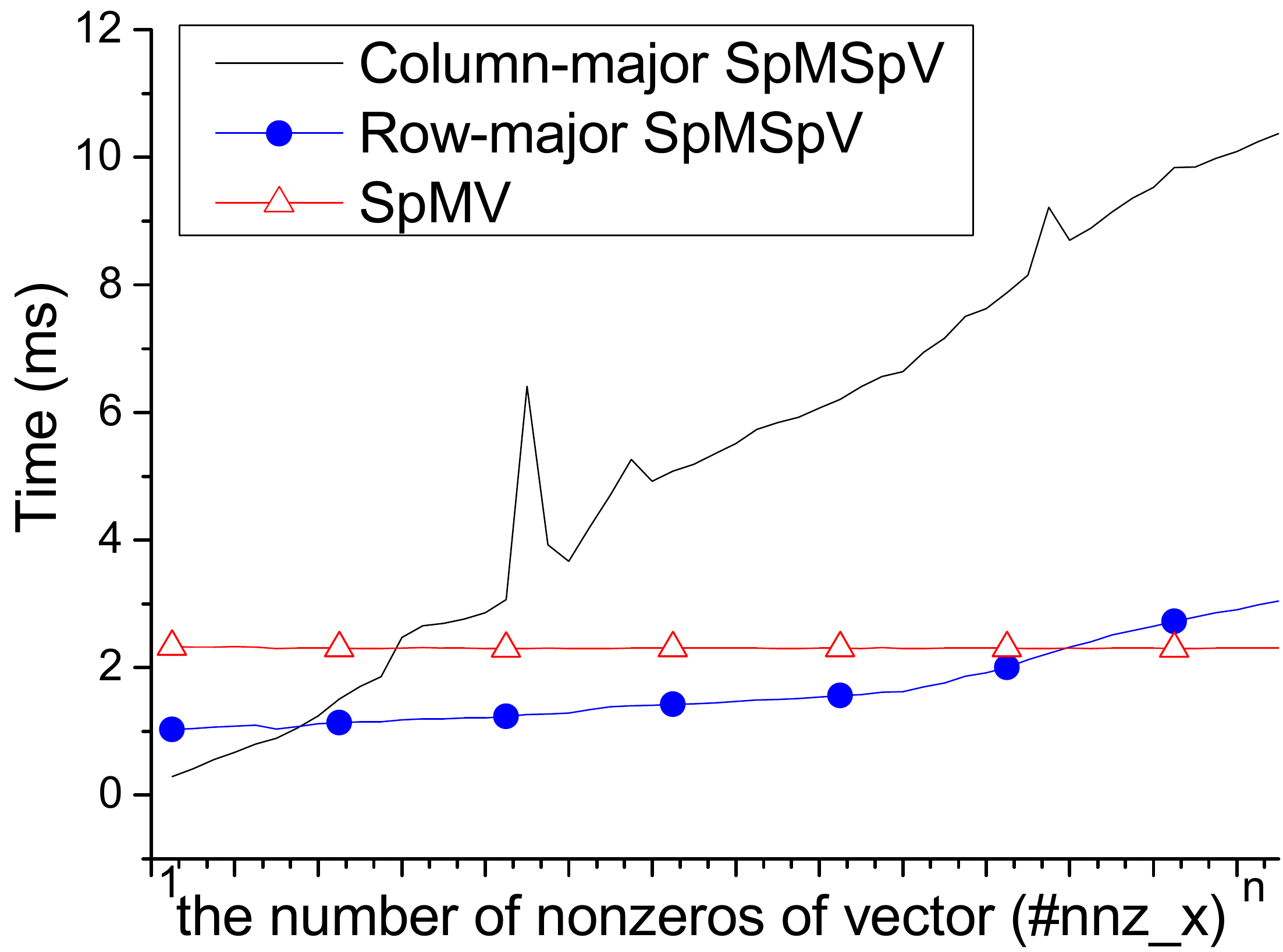}}
\caption{The performance of SpMV, row-major and column-major SpMSpV solutions with input vectors of various sparsity: (a) \textit{ljournal-2008} (b) \textit{wikipedia-2007}.  }
\label{fig-computation-pattern}
\end{figure}

To further verify that these three SpMV/SpMSpV solutions are feasible choices
for different cases, we conduct an experiment to examine their performance
with randomly generated inputs vectors, whose number
of nonzeros vary from 1 to $n$ with an uniform interval.
The data sets tested are \textit{ljournal-2008} and \textit{wikipedia-2007}.
The test results are shown in Figure~\ref{fig-computation-pattern},
where we take the SpMV from \cite{hola-spmv} and the column-major SpMSpV
from \cite{spmspv-sort-2015}. The row-major SpMSpV is implemented by
integrating the SpMV kernel with simple value validation operations.
From the test results we can see that column-major SpMSpV is more promising
when the input vector is very sparse, the traditional SpMV is of no doubt
the suitable choice when the input vector is very dense, and in between of
them the row-major SpMSpV is potentially more efficient.
This further implies the necessity of designing an adaptive framework.

\subsection{Implementation Analysis}
\label{parallel-imp}

The implementation of the candidate SpMV/SpMSpV solutions mainly involves two aspects.
One is how to distribute the workload to the different threads.
The other is how to write back the result to the output vector.
Different workload distribution and write-back strategies may have a strong influence on the performance.
We will take these two factors into consideration so that the kernel selection space can be further expanded.

\subsubsection{Workload distribution}
\label{workload}

\begin{figure}[!htb]
\centering
\includegraphics[width=0.97\columnwidth]{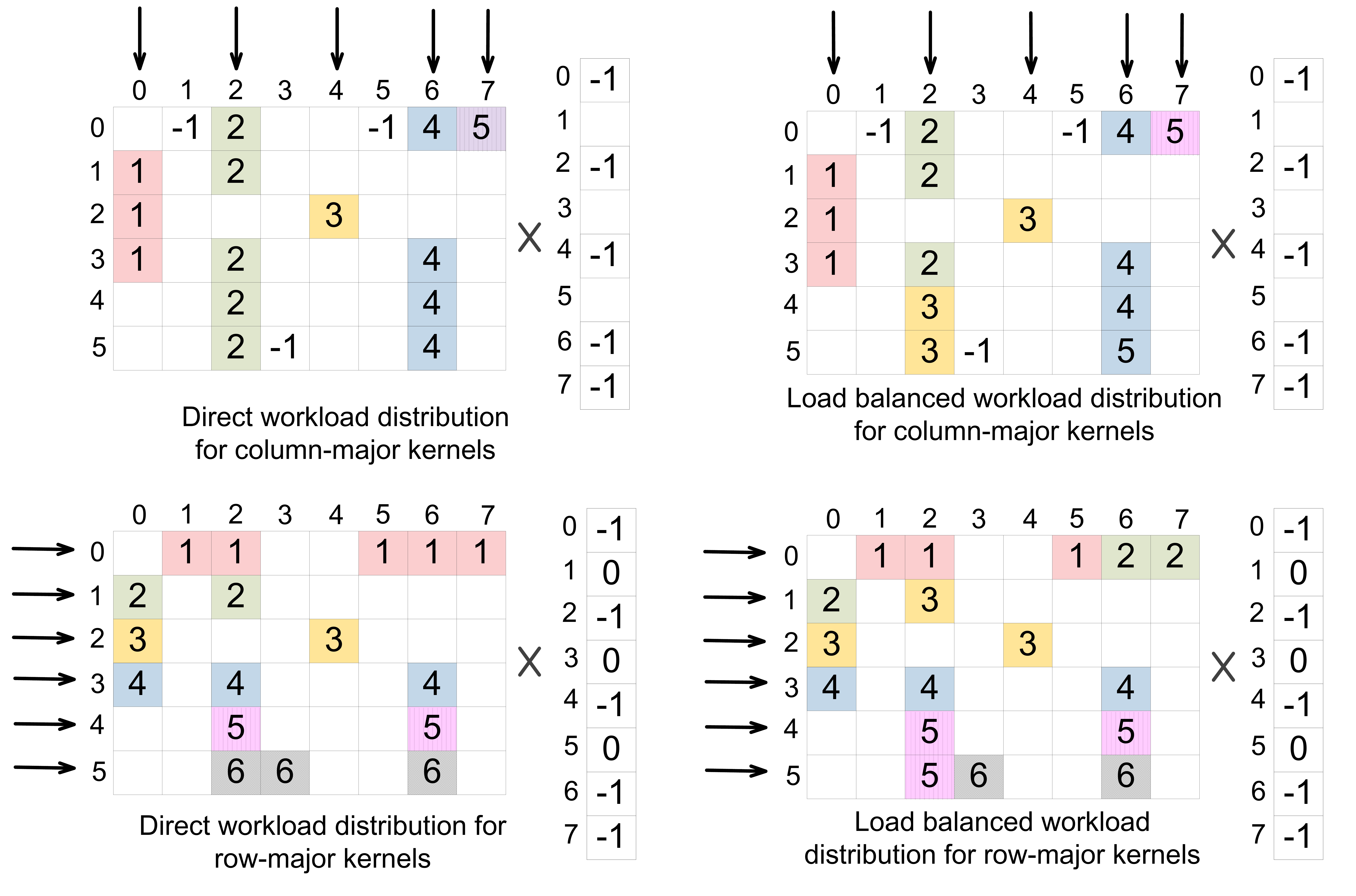}
\caption{The illustration of different work distribution methods for column- and row-major SpMSpV kernels.}
\label{fig-work-distribution}
\end{figure}

In the implementation of SpMV and SpMSpVs kernels, the major workload is to access the nonzeros of the matrix.
There are two types of workload distribution methods.
The first, straightforward as it is, distributes a column or a row of the matrix to each thread.
The other method relies on some preprocessing strategy \cite{Merge-spmv,hola-spmv}
to help partition the total workload, i.e., nonzero entries, evenly to each thread.
The two work-distribution methods for both column and row-major SpMSpV
are illustrated in Figure~\ref{fig-work-distribution},
where the workload for different threads are shown by different colored numbers,
and the matrix access directions are indicated by  arrows.

\begin{figure}[!t]
\centering
\subfigure[\textit{hugetrace-00020}]{\includegraphics[width=0.33\columnwidth]{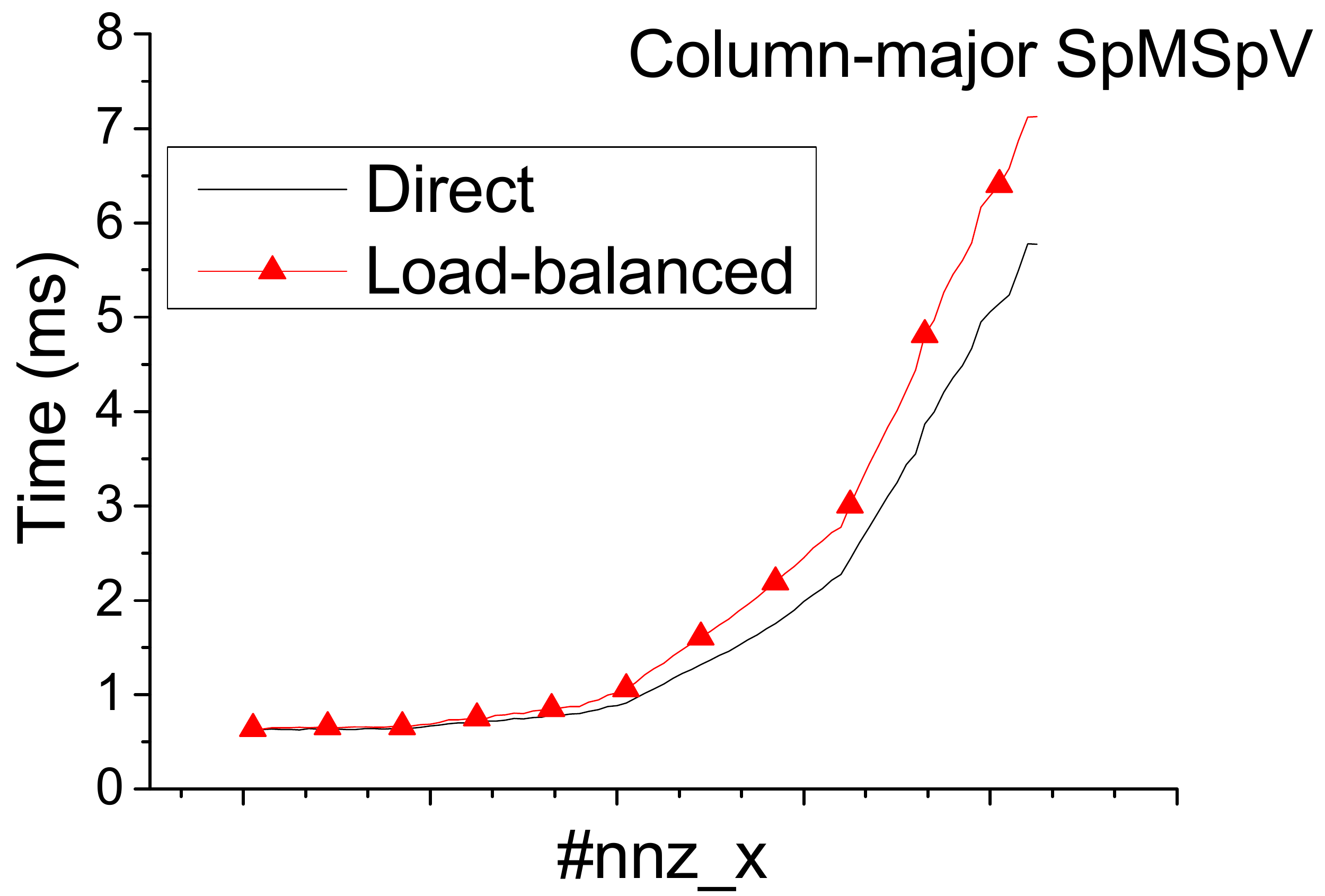}
\includegraphics[width=0.32\columnwidth]{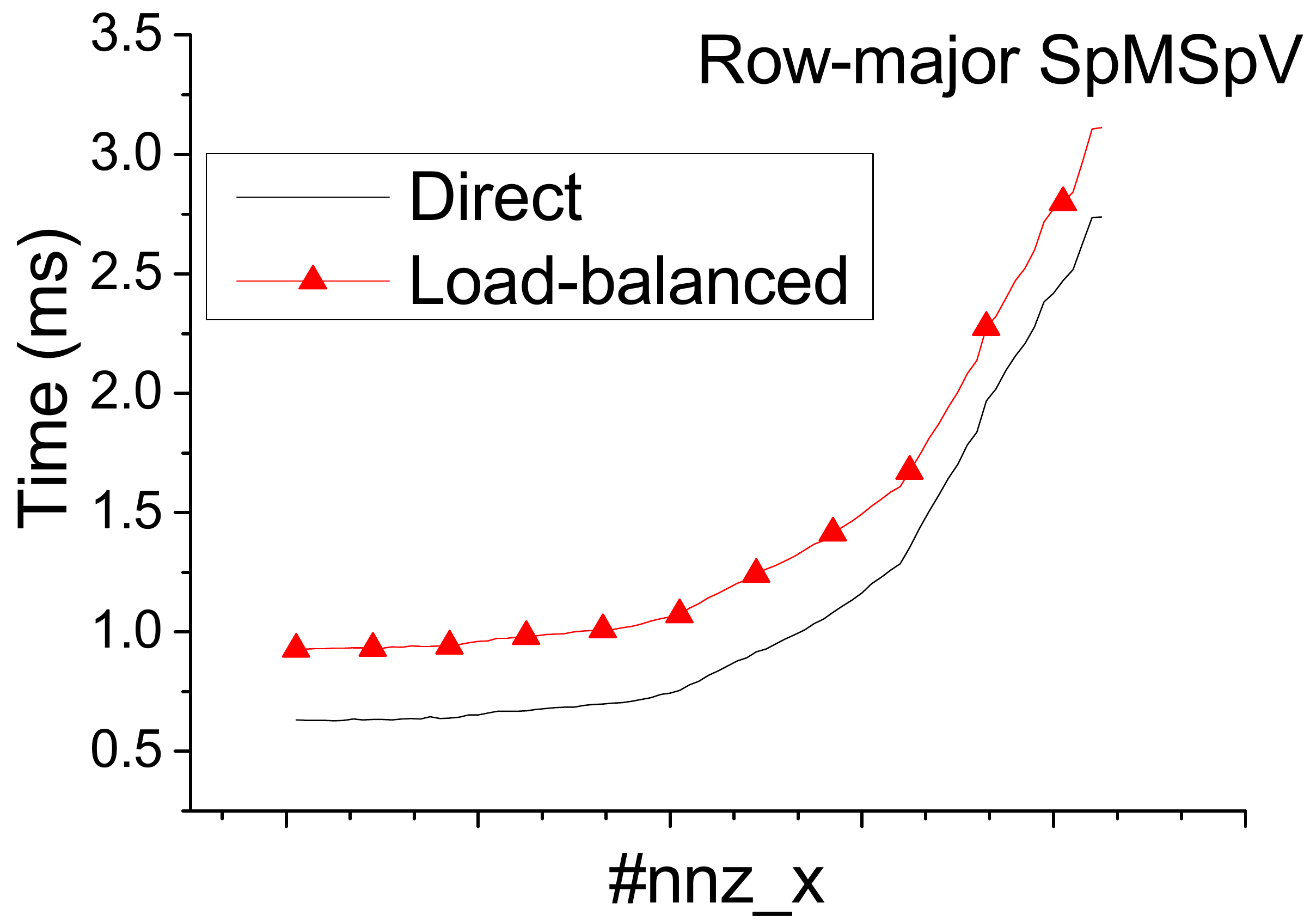}
\includegraphics[width=0.3\columnwidth]{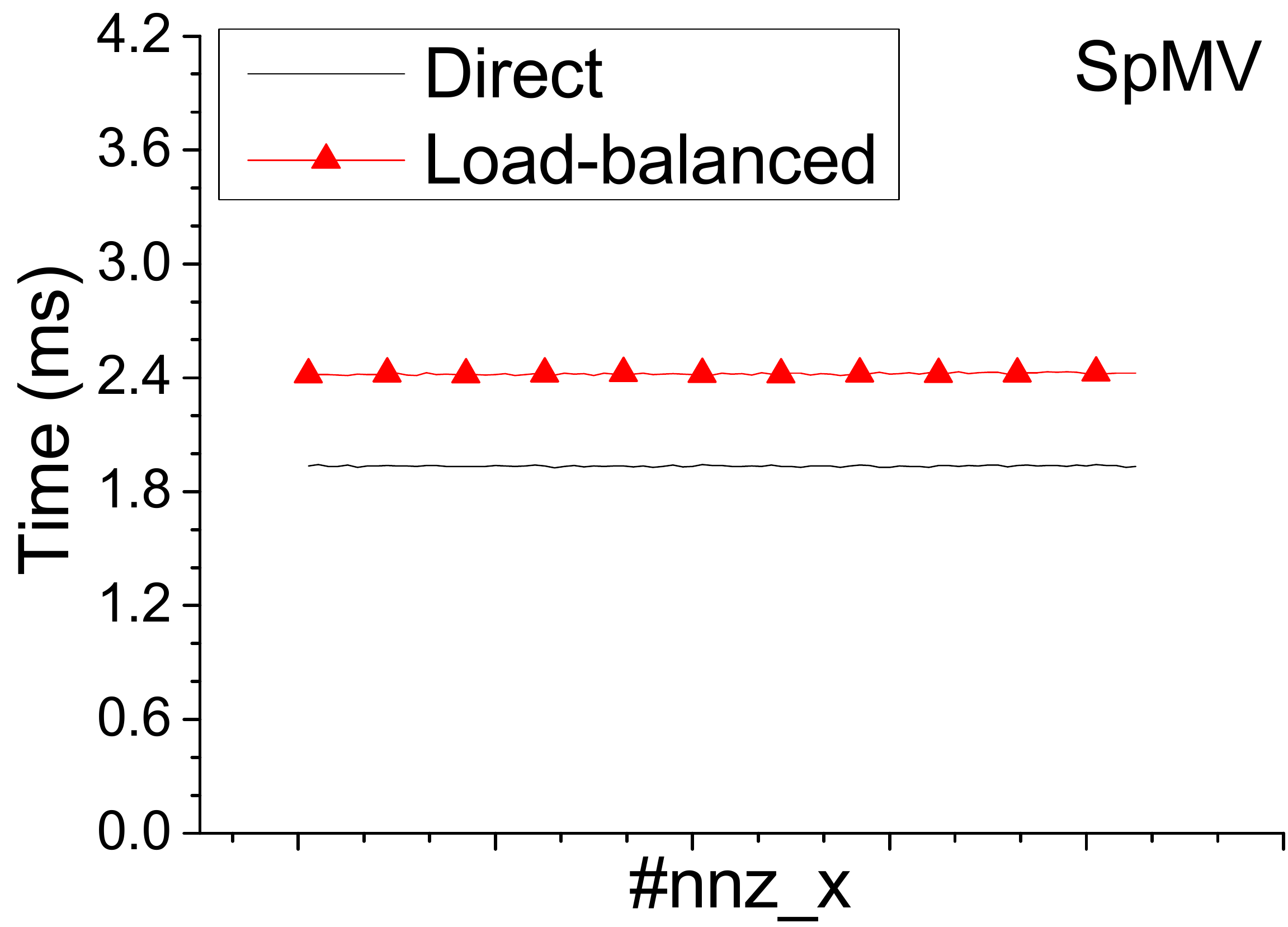}}
\subfigure[\textit{wikipedia-2007}]{\includegraphics[width=0.33\columnwidth]{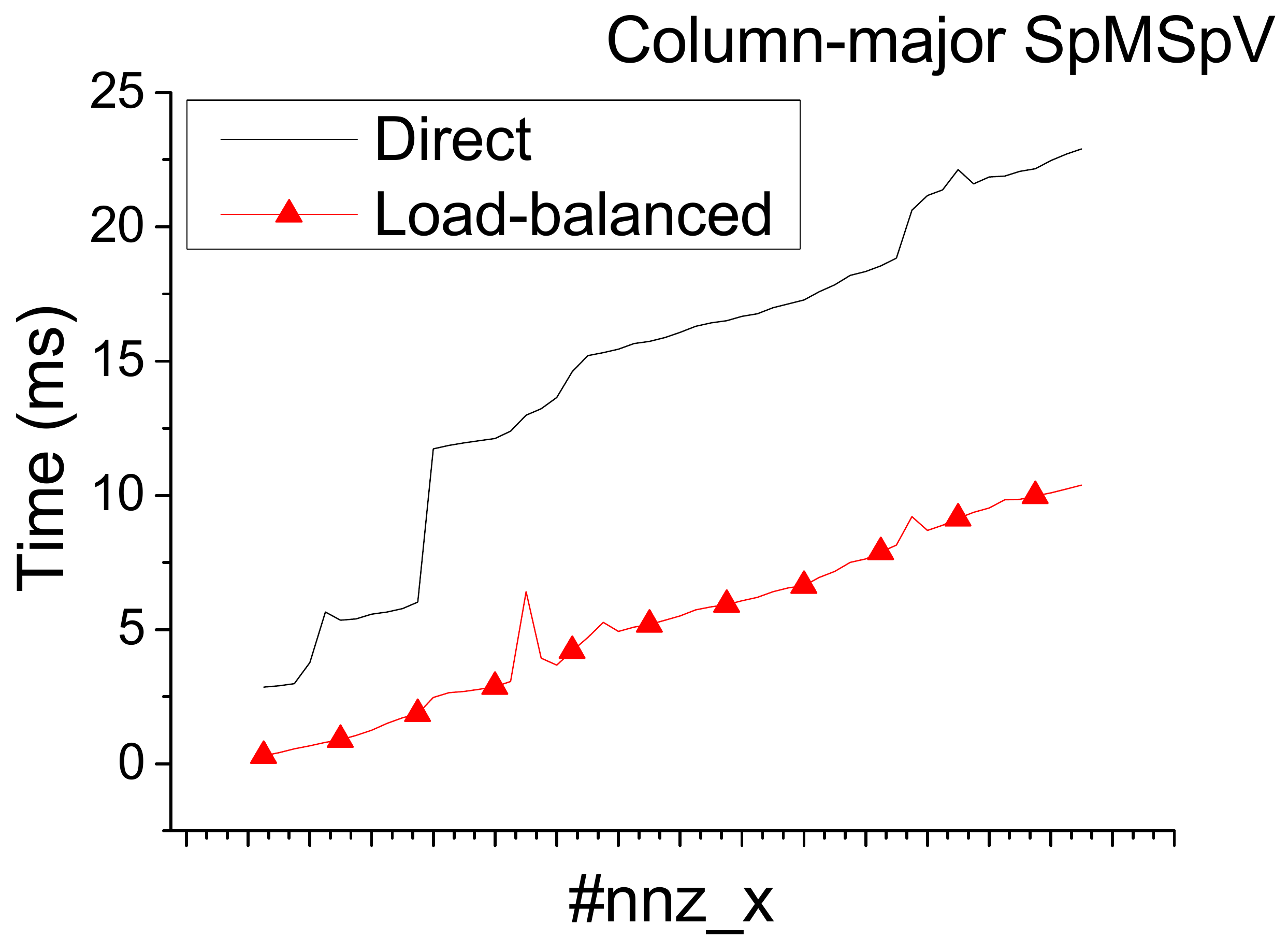}
\includegraphics[width=0.32\columnwidth]{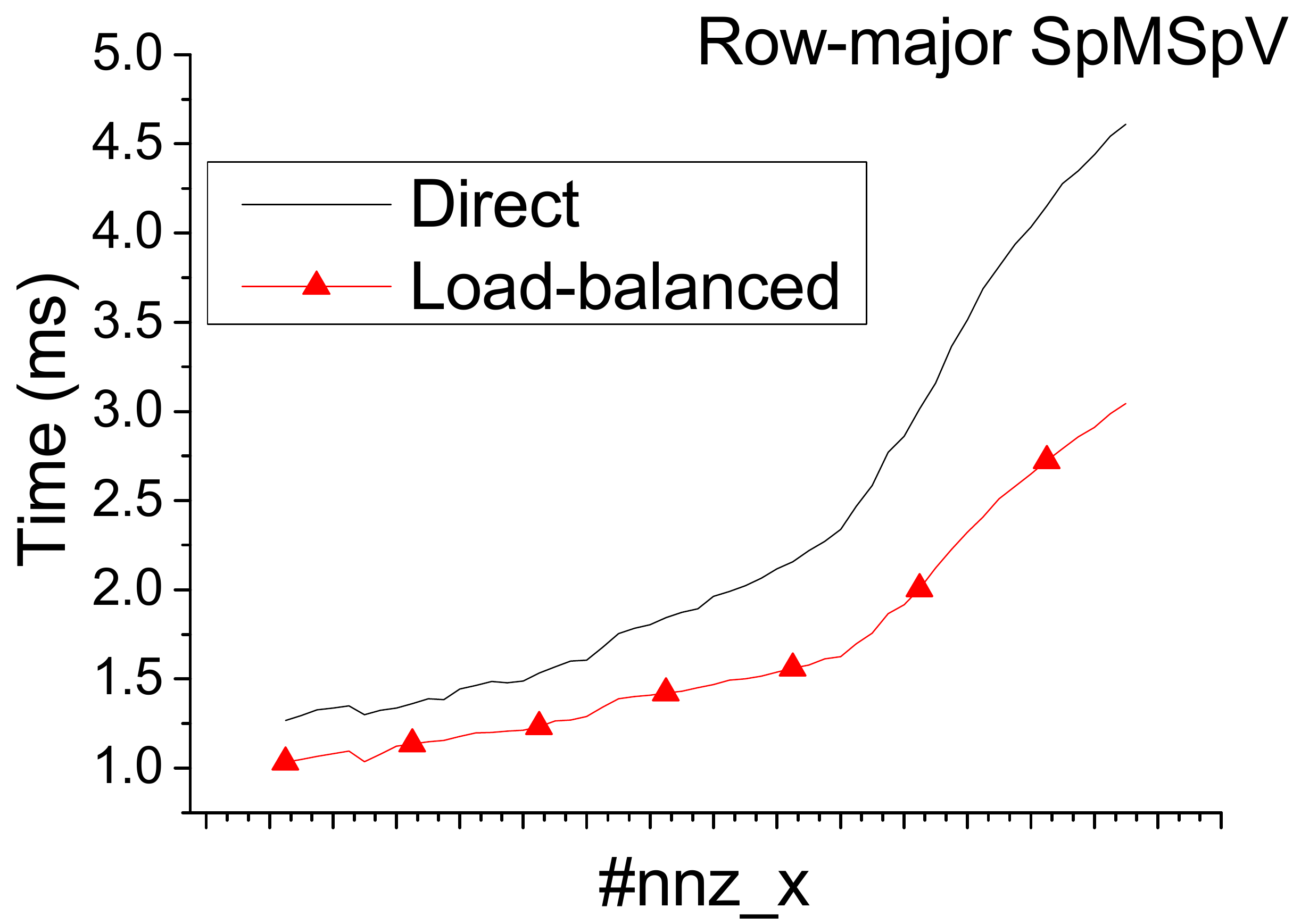}
\includegraphics[width=0.3\columnwidth]{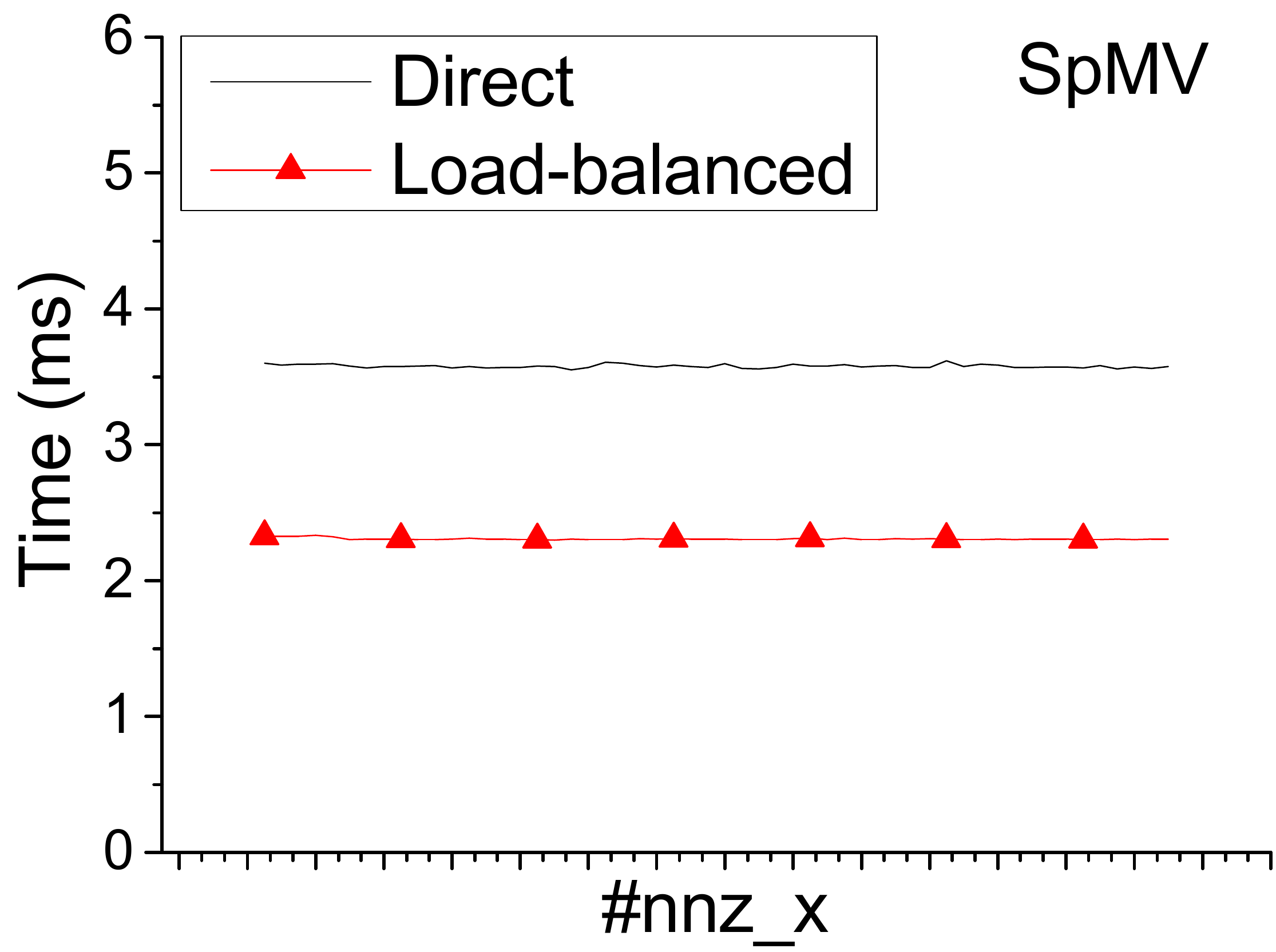}}
\caption{The performance of the two workload distribution methods for the candidate solutions: (a) \textit{hugetrace-00020}, (b) \textit{wikipedia-2007}.}
\label{fig-load-balance}
\end{figure}

To examine the performance of the two work distribution strategies,
we conduct an experiment on the three candidate SpMV/SpMSpV solutions,
with the number of nonzeros of the inputs vectors varying from 1 to $n$ with an uniform interval.
The data sets we use are \textit{hugetrace-00020}, which has a relatively uniform distribution of nonzeros,
and \textit{wikipedia-2007}, which has a strong variation in the nonzero distribution.
From the test results shown in Figure~\ref{fig-load-balance},
we can observe that the performance of the two work distribution {methods} are very different
for the two data sets. When the nonzero entries of the matrix is  uniformly  spread,
the direct method is more efficient, while when the nonzero distribution is non-uniform,
the load-balanced is more promising.

\subsubsection{Write-back}
\label{writing-back}

In addition to workload distribution, the write-back strategy is also important for SpMSpV computations.
With the required data accessed and loaded from the matrix, each thread needs to
multiply them with the vector values
and write the partial sum back to the corresponding location in the output vector.
For row-major SpMSpV, the matrix is accessed along the row, which is consistent with the write-back order,
thus incurring little write conflict.
If a column-major solution is applied, the matrix is accessed along the column, which no longer
matches with the write-back order. In this case, race conditions would occur
when different threads try to write to the same memory location.
It is therefore necessary to consider different write-back strategies
for the column-major SpMSpV and further expand the {kernel} space.

In this paper, two types of write-back strategies are taken into considerations.
One utilizes atomic operations to write back directly, which is achieved by either hardware support
or software simulation.
For this method, the (row\_id, value) pairs for effective nonzeros of the matrix do not need to
write back to the global memory and only one kernel function is spawn to finish all the works.
The other approach is based on sort \cite{spmspv-sort-2015} and requires to spawn three kernel functions.
It first writes the (row\_id, value) pairs for effective nonzeros of the matrix to the global memory,
then spawns a kernel to sort the (row\_id, value) pairs,
and finally the reduce-by-key primitive is used to write back results without conflict.
This sort-based method can effectively get rid of the write conflict,
but it also brings redundant global memory accesses.
Without an adaptive framework, it is hard to tell which method is superior.

\begin{figure}[!htb]
\centering
\subfigure[Direct]{\includegraphics[width=0.49\columnwidth]{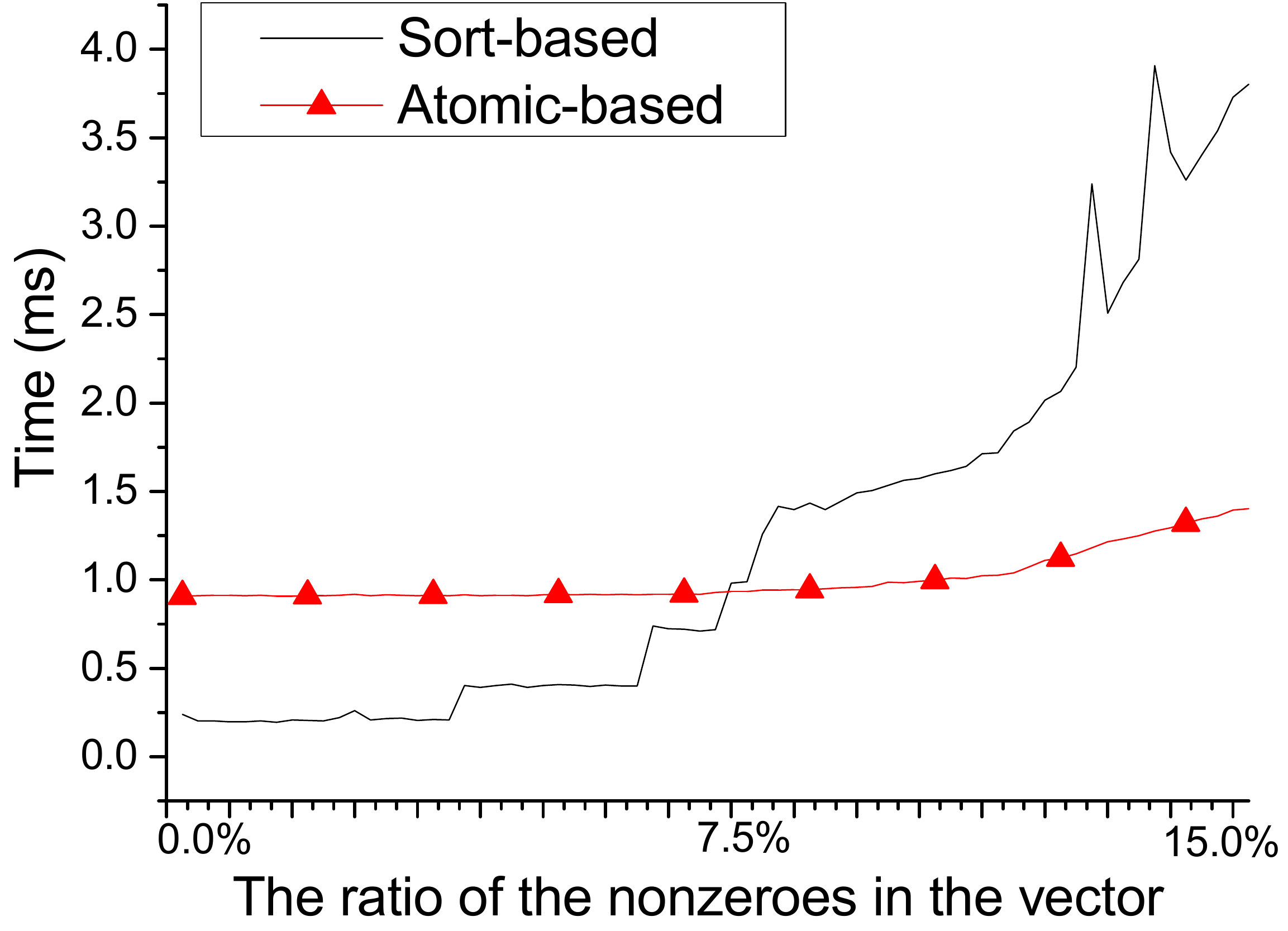}}
\subfigure[Load-balanced]{\includegraphics[width=0.49\columnwidth]{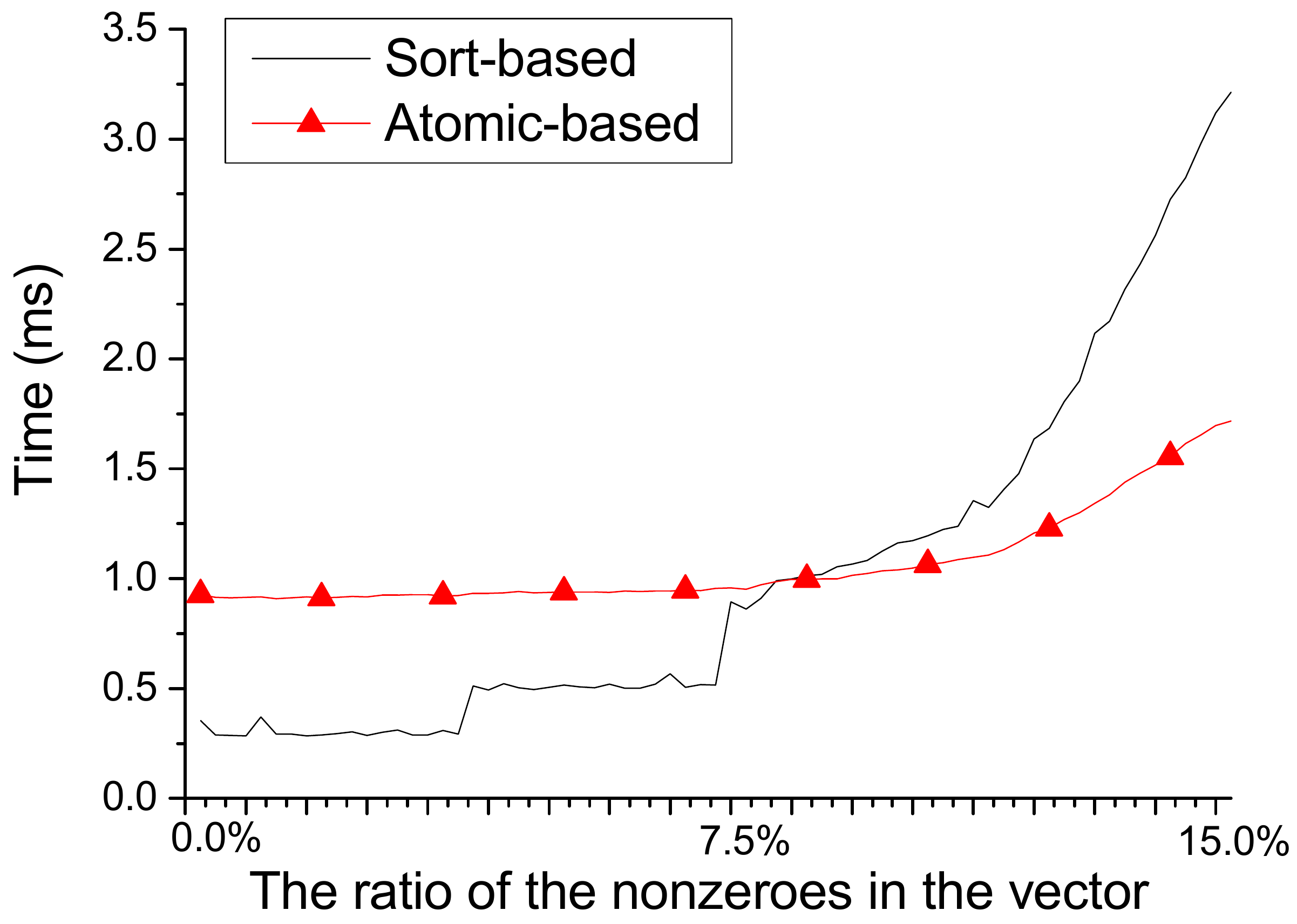}}
\caption{The performance of the two write-back strategies on the \textit{road\_usa} data sets for both
(a) direct column-major SpMSpV and (b) load balanced column-major SpMSpV.}
\label{fig-writing-back}
\end{figure}

We carry out an experiment on the column-major SpMSpV to examine the performance of
the two write-back strategies.
The \textit{road\_usa} data set is taken as an example and both the direct and load-balanced
versions of the column-major SpMSpV are tested.
The test results are drawn in Figure~\ref{fig-writing-back}.
From the figure, it can be seen that the sort-based method can play a role
when the vector is relatively sparse but starts to lose the advantage
as the vector becomes denser.

\section{Kernel Selector}
\label{model}

In the previous section, we devise eight candidate kernels with considerations
made on key factors including the computing pattern,
workload distribution and write-back strategies.
It is then of interest to find a way to select the most suitable kernel that can sustain
the best performance to adapt with the varieties of both input data and hardware architectures.
This is done by transforming the kernel selection process into a series of classification problems,
which are solved by using machine learning models in this work.
In particular, to select the appropriate SpMV/SpMSpV kernel, several classification problems
are solved, for which three machine learning models
are trained and utilized to make predictions based on the computing pattern,
workload distribution and write-back strategies, respectively.
Once the models are trained, one can use them to make the prediction on-the-fly as many times as necessary.

The whole kernel selection process is comprised of two major steps:
feature extraction and model generation, which will be detailed as follows.

\tabcolsep 1.4pt
\begin{table}[!h]
\scriptsize
\footnotesize
\renewcommand{\arraystretch}{1.3}
\caption{Extracted features for the three machine learning models.}
\label{table:features}
\centering
\begin{tabular}{|c|l|c|}
 \hline
Features & Meaning  &Model\\
 \hline
   $m$                &  number of rows                     & 1,2,3 \\
   $n$                &  number of columns                  & 1,2,3 \\
   $nnz$              &  number of nonzeros                & 1,2,3   \\
   \hline
   $\mathrm{max\_row}$           &  maximum numbers of nonzeros per row               &1,2\\
   $\mathrm{min\_row}$           &  minimum numbers of nonzeros per row               &1,2\\
   $\mathrm{avg\_row}$           &  average numbers of nonzeros per row               &1,2\\
   $\mathrm{relative\_range}$    & $(\mathrm{max\_row}-\mathrm{min\_row})$/$n$             &1,2\\
   $\mathrm{var\_nnz\_row}$      &  standard variation of nonzeros per row  &1,2\\
   $\mathrm{G_c}$                 &  Gini coefficient \cite{gini}                            &1,2 \\
 \hline
   $nnz\_x$                &number of nonzeros of the vector                &1,3 \\
   $x\_\mathrm{sparsity}$  &ratio of nonzeros of the vector         &1,3 \\
   $nnz\_s$                &number of effective nonzeros of the matrix      &1,3 \\
   $m\_\mathrm{sparsity}$  &ratio of effective nonzeros of the matrix    &1,3 \\
   \hline
  \end{tabular}
\end{table}

\begin{figure*}[!htb]
\centering
\subfigure[Computing pattern]{\includegraphics[width=0.66\columnwidth]{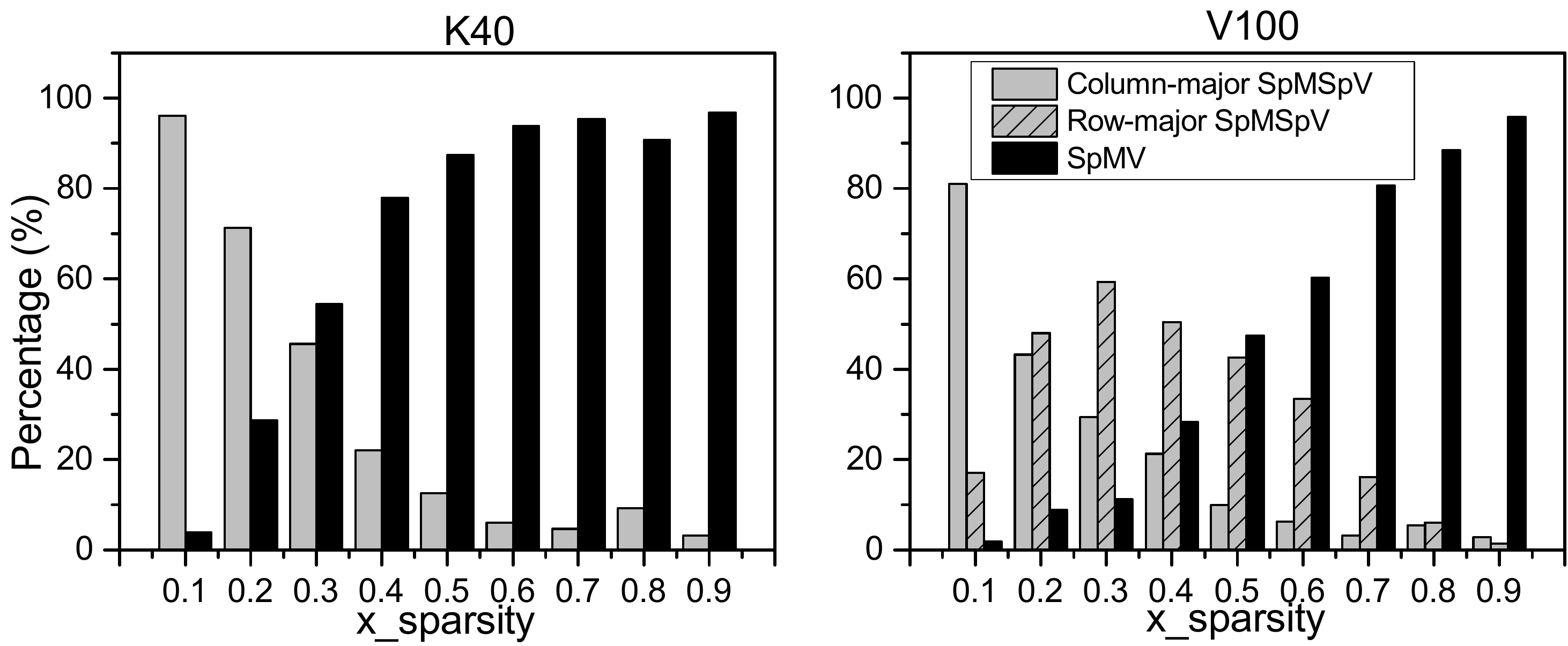}}
\subfigure[Workload distribution]{\includegraphics[width=0.66\columnwidth]{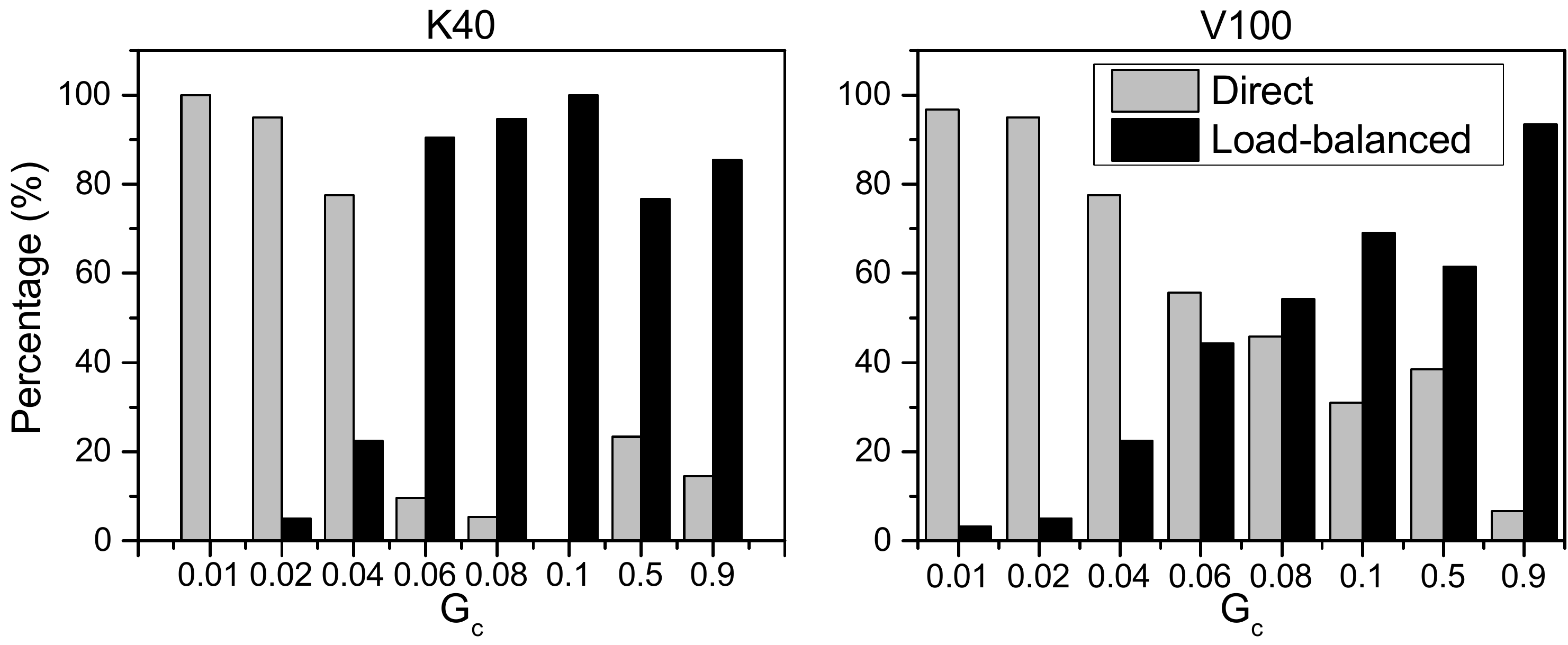}}
\subfigure[Write-back]{\includegraphics[width=0.66\columnwidth]{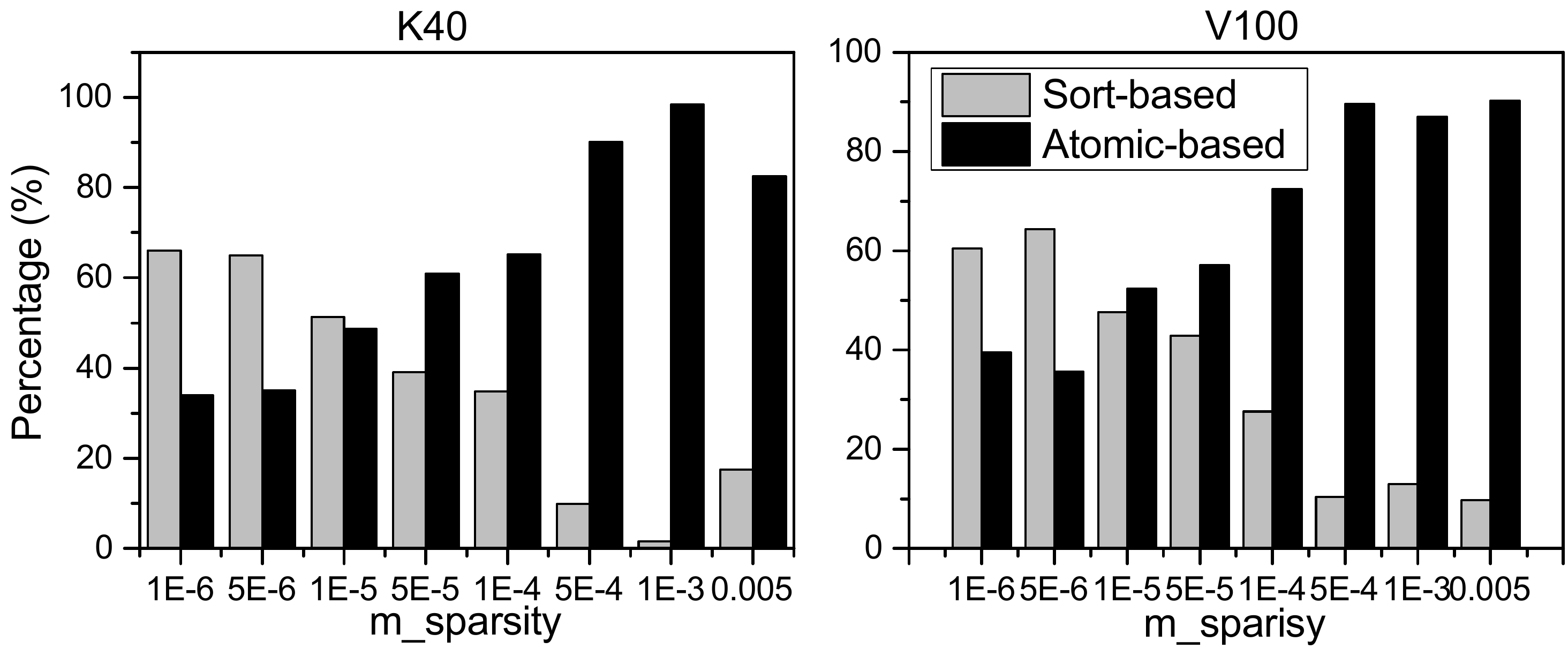}}
\caption{The statistical information of the samples under the influence of one of the most important features,
where y-axis shows the percentage of samples falling into each parameter value interval.}
\label{fig-features}
\end{figure*}

\subsection{Feature Extraction}
For the machine learning models, the feature space is designed based on
the collection of a variety of features listed in Table~\ref{table:features},
in which we also show the correspondence between the selected features and the three models.
{For the convenience of expression, the models for computing pattern, workload distribution,
and write-back are denoted with 1, 2 and 3 in the table, respectively.}
Among these features, parameters related to the dimension of the input matrix,
i.e., the numbers of rows, columns, and nonzeros are extracted
to represent the basic information of the matrix to be used by all three models.
A number of specific features are chosen to describe different behaviors
of the selection alternatives.
In terms of workload distribution, the sustained performance could be greatly
influenced by the distribution of the nonzeros of the input matrix.
Therefore, parameters that can measure the distribution of nonzeros are extracted,
such as the maximum, minimum and average numbers of nonzeros per row,
the standard variation of nonzeros per row,
the maximum and minimum ratios of nonzeros per row,
and the Gini coefficient \cite{gini}.
The Gini coefficient is a measure of the equality of degree distributions for graphs,
which can also be used to measure the distribution of nonzeros of a sparse matrix.
As for the selection of write-back strategies,
the performance could have a strong correlation with atomic operations
and redundant memory accesses.
Therefore, to measure them we extract features such as
the number and ratio of nonzeros of the input vector, i.e., $nnz\_x$ and $nnz\_x/n$,
as well as the number and ratio of the effective nonzeros of the input matrix, i.e., $nnz\_s$ and $nnz\_s/nnz$.
With respect to different computing patterns, all the aforementioned features
are potential factors and are included in the corresponding machine learning model.
To see the importance of these features \cite{ada-trsv}, the \texttt{SelectKBest} method in scikit-learn package
with chi-squared score function is used.
For the three models, the feature sequences according to their importance are \{$nnz$, $nnz\_s$, $n$, $m$, $nnz\_x$, $x\_\mathrm{sparsity}$,
$m\_\mathrm{sparsity}$, $\mathrm{G_c}$, $\mathrm{max\_row}$, var\_nnz\_row, $\mathrm{avg\_row}$, $\mathrm{min\_row}$, $\mathrm{relative\_range}$ \},
\{$nnz$, $m$, $n$, $\mathrm{G_c}$, $\mathrm{max\_row}$, var\_nnz\_row, $\mathrm{avg\_row}$, $\mathrm{min\_row}$, $\mathrm{relative\_range}$\}
and \{$nnz$, $m$, $n$, $nnz\_s$, $nnz\_x$, $m\_\mathrm{sparsity}$, $x\_\mathrm{sparsity}$ \}, respectively.

We also show some statistical information on different architectures.
Figure \ref{fig-features} plots the distribution of samples under the influence of one of the most important features of the three models,
where y-axis shows the percentage of samples falling into each parameter value interval.
For convenience of expression, NVIDIA Tesla K40 and V100 GPUs are used as representatives to present the results.
As seen from Figure \ref{fig-features}, the ratios of the samples falling into each interval are different on different hardware platforms
in terms of computing pattern, workload distribution and write-back methods. That means the learned rules or the switch thresholds of the rules
can be different from one hardware architecture to another.
Thanks to the use of machine learning methods, the adaptive framework can be flexibly adapted to different hardware platforms.

\subsection{Model Generation}

To construct the data set for training the machine learning models,
2,010 sparse matrices
are taken from the extensively utilized SuiteSparse matrix collection \cite{matrix-market},
which covers a large variety of application fields and can serve as the representative of
different matrix distributions.
For each sparse matrix, we test many groups of SpMV/SpMSpV computations
by gradually increasing the nonzeros of input vector from $1$ to $n$ with randomly generated entries.
The performance of all the candidate kernels are collected
and data samples containing the aforementioned features and a unique label are generated.
The label is determined by comparing the performance of the corresponding selection alternatives.
The data samples are then divided into two data sets with a split ratio of 7:3, among which the former is used for training
and the latter for testing.
The machine learning models are trained with the training data and integrated into the adaptive framework.
Once the machine learning models are generated, they can be used to make the prediction on-the-fly as many times as necessary.

To setup the machine learning models, we explore a number of different choices,
including decision tree, SVM, random forest and gradient tree boosting (GBDT).
All of the models are implemented based on the scikit-learn package \cite{sklearn}.
To tune the hyper-parameters in each model, the grid search technique with cross-validation is utilized
to perform an exhaustive search over a range of parameters and find the best parameter set.
For decision tree, we explore the parameters such as the maximum depth of the tree, i.e, \texttt{max\_depth},
which is set in the range of [1,10], and the class weights, which belong to \{`balanced', `uniform'\}.
For SVM, the RBF kernel is utilized, with the regularization parameter
in the range of [1, 1000] and the kernel coefficient belonging to \{`auto', `scale'\}.
For random forest, we set the number of trees to be used in the forest
to be among \{50, 100, 200, 500\}
and the maximum depth of the tree \texttt{max\_depth} to be in the range of [1, 10].
As for GBDT, apart from the same setup for \texttt{max\_depth},
the learning rate is set within \{0.1, 0.01\} and the number of boosting stages are set
to be among \{50, 100, 200, 500\}.

\tabcolsep 1.4pt
\begin{table}[!h]
\scriptsize
\footnotesize
\renewcommand{\arraystretch}{1.3}
\caption{Classification accuracy and overhead of the models.}
\label{table:accu}
\centering
\begin{tabular}{|c|c|c|c|c|l|l|l|}
 \hline
  \multirow{2}*{Methods} &\multirow{2}*{Model}  & \multicolumn{3}{c|}{Accuracy} & \multicolumn{3}{c|}{Overhead}\\  \cline{3-8}
                     &   & K40   & P100  & V100 & K40  & P100  & V100 \\   \cline{1-5}\cline{1-8}
  \multirow{4}*{\shortstack{Computing \\pattern}}
         &Decision tree    &93.9\%   &89.9\%   &90.3\%  & 1.0x  & 1.0x &1.0x    \\ \cline{2-8}
         & SVM             &-0.03\%  &-2.3\%   &-5.4\%  & 14.3x & 21.7x & 22.8x \\   \cline{2-8}
         & Random forest   &+2.6\%   &+1.4\%   &+2.5\%  & 71.1x & 83.1x & 83.0x \\   \cline{2-8}
         & GBDT            &+2.6\%   &+2.4\%   &+2.9\%  &7.1x  &30.1x &31.4x    \\   \cline{1-8}\cline{1-8}

  \multirow{4}*{\shortstack{Workload \\distribution}}
         & Decision tree      &95.3\%   &97.9\%    &98.7\%  &1.0x   &1.0x & 1.0x   \\  \cline{2-8}
         & SVM                &+3.3\%   &+1.1\%    &+0.3\%  &21.5x  &35.8x & 39.2x       \\  \cline{2-8}
         & Random forest      &+3.3\%   &+0.8\%    &+0.0\%  &42.2x  &44.1x & 43.3x   \\   \cline{2-8}
         & GBDT               &+3.6\%   &+1.1\%    &+0.3\%  &8.3x   &7.9x &8.0x    \\  \cline{1-8}\cline{1-8}
   \multirow{4}*{Write-back}
         & Decision tree      &92.7\%  &96.8\%     &97.2\%  &1.0x  &1.0x   &1.0x   \\ \cline{2-8}
         & SVM                &+0.5\%  &-8.7\%     &+11.8\% &21.0x &4.2x   &3.3x       \\ \cline{2-8}
         & Random forest      &+4.5\%  &+1.7\%     &+2.5\%  &81.9x &6.0x   &6.3x       \\ \cline{2-8}
         & GBDT               &+4.7\%  &+0.2\%     &+2.6\%  &8.8x  &4.7x   &4.2x   \\ \cline{1-6}\cline{1-8}
   \hline
  \end{tabular}
\end{table}

To find out the most appropriate machine learning model for the kernel selector, we compare the performance of the aforementioned models
by experiments. Results with the test data in terms of both accuracy and overhead are summarized in Table \ref{table:accu},
in which the accuracy is shown as the relative difference with the decision tree model
and the overhead is calculated as the normalized factor with respect to the decision tree model.
It can be seen from the table, all tested machine learning models can achieve relatively high accuracy,
ranging from 84.9\% to 99.0\%. This further validates the effectiveness of the extracted features.
In addition, when compared with decision tree, other machine learning models may
have better accuracy, though the improvement is relatively small.
However, the prediction overhead for these more accurate models are substantially
larger than that of the decision tree model.
For example, the GBDT model is around 0\% to 4.7\% more accurate than decision tree,
but its overhead is 4.2x to 31.4x higher,
which could make the adaptive framework lose its performance superiority.
Based on the test results, users can select the appropriate models according to the specific requirement
of accuracy and prediction overhead. In this paper, the decision tree models are incorporated
as a demonstration to present performance of the proposed adaptive framework.

\section{Runtime Procedure}
\label{runtime-process}

To select the proper candidate from the eight SpMV/SpMSpV kernels
by using the decision tree models,
some further details are also of importance to the performance
of the adaptive framework. These include the implementation of the candidate kernels,
the execution procedure of the models, and the optimization of models for prediction.

\subsection{Kernel Implementation}

Integrated into the adaptive SpMV/SpMSpV framework are overall eight candidate kernels,
including two SpMV kernels, two row-major SpMSpV kernels and four column-major SpMSpV kernels.
Each of the eight candidates is implemented as a standalone GPU kernel.
The implementation of the two SpMV kernels is based on the work of \cite{naive-spmv} and
the latest Hola-SpMV library \cite{hola-spmv}, corresponding to the
direct and load-balanced versions, respectively.
Since the computation skeleton of row-major SpMSpV is very close to SpMV except the value validation part,
one can make use of the existing SpMV implementations by adding value validation into the code
instead of implementing the whole kernel from scratch.
The bitvector technique presented in \cite{GraphMat-2015} is used to accelerate the value validation process
by using bit operations.
As for column-major SpMSpV, we have four versions considering different workload distribution and write-back strategies.
The load-balanced and the sort-based implementations can be extracted from \cite{spmspv-sort-2015}.
The direct implementation can be easily acquired by modifying the direct implementation in SpMV \cite{naive-spmv}
by assigning the computation of one column of the matrix to each thread.
And the atomic version can be directly implemented by using  atomic instructions.
We remark that users are free to replace any of the eight kernels with other high performance implementations
into the framework for their own needs.

To encapsulate the aforementioned SpMV/SpMSpV kernels into the adaptive framework, seamless
transitions among different data formats of the input matrix and vector are needed during
the program execution. Although the sparse matrix keeps unchanged, the overhead of the format
transformation can still be high. In order to reduce the cost, we store both the CSR and CSC formats
for the sparse matrix at the beginning.
Similar approaches have been adopted in previous works such as \cite{direction-bfs, spmspv-push-pull-2018}.
The input vector, which is provided by the application every
iteration, is converted to the selected vector format on-the-fly as needed. And the cost of this online
vector transformation, as will be shown later by experiment results, is relatively small.

\subsection{Model Execution}

\begin{figure}[!t]
\centering
\includegraphics[width=0.98\columnwidth]{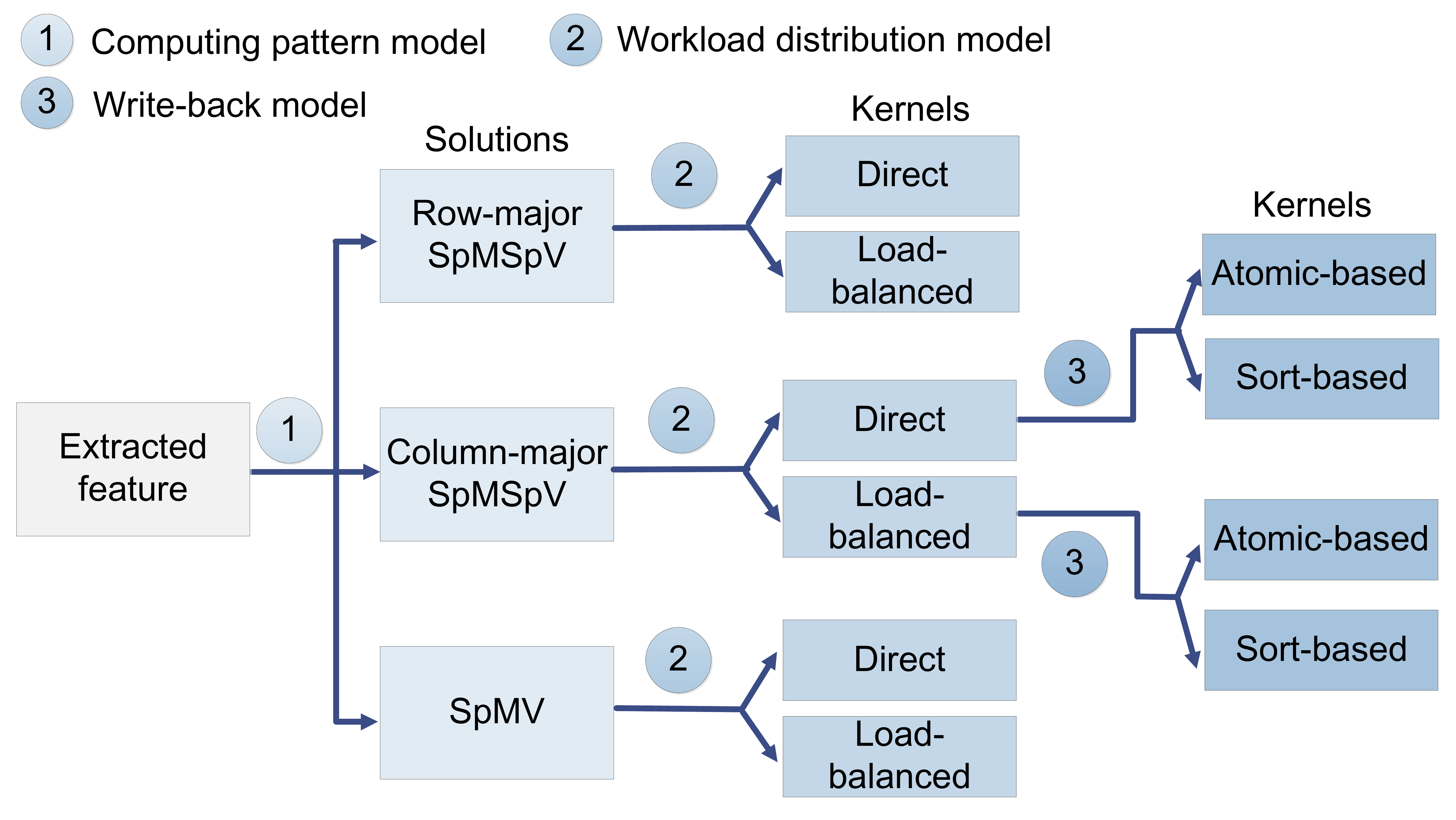}
\caption{The model execution procedure of the three decision tree models.  }
\label{fig-runtime}
\end{figure}

After training, the three decision tree models are utilized for inference at the runtime.
Figure~\ref{fig-runtime} shows the execution process of the three models,
during which the rules for different models are checked one by one in a cascaded fashion.
The model for computing pattern selection is first executed
and one of the candidate solutions, i.e., column-major SpMSpV, row-major SpMSpV, and SpMV, are selected.
Based on this, the workload-distribution model is utilized to predict which method, direct or load-balanced, is more suitable.
If the column-major SpMSpV is selected in the first step, the model for write-back strategy is further employed to
chose whether atomic-based or sorted-based strategy is used to write back the result.
It is worth noting that the order of execution of the workload-distribution and write-back models
is commutable for column-major SpMSpV, which does not affect the final results.

\subsection{Runtime Optimization}
In the runtime prediction process, the major cost includes the computation of
features, the evaluation of the decision tree rules,
and if necessary, the conversion of the vector format.
Among them, the matrix related features are computed only once at the beginning
of the framework, and can be reused across different iterations.
The vector related features and the decision tree rules, however, are required to
be computed at every iteration. If and only if the predicted solution
in current iteration is not the same as the previous one, the vector format conversion
is triggered.

In order to reduce the overhead of the runtime prediction, we utilize several optimization techniques. First, the processes of both the feature extraction and vector format conversion are parallelized on the GPU by utilizing well-tuned parallel primitives, such as reduction, prefix sum and stream compaction.
Second, the decision tree models are further optimized to accelerate the kernel selection process.
Particularly, those rules with lower feature computation overhead, such as rules including $m, n, nnz, nnz\_x$ and $x\_\rm{sparsity}$, will be given higher priority to perform.
In addition, instead of computing all the features ahead of time, the evaluation of each feature is lazily done
only when the feature is used, which is potentially helpful for reducing the execution overhead.
For example, if an input case can be classified by using the rule including feature $n$,
then there is no need to evaluate other features.
Our experiments show that these optimizations can improve the performance of the runtime execution
by around 2x - 17x.

\section{Experiment Results}
\label{experiment}

\tabcolsep 3.2pt
\begin{table}[!tb]
\footnotesize
\caption{Overview of the data sets for experiments.}
\label{table:dataset}
\centering
\begin{tabular}{|c|c|r|r|c|}
\hline
Data sets&Spyplot  &$n$ &$nnz$  &Abbr. \\
\hline
belgium\_osm &\mgape{\includegraphics[scale=0.025]{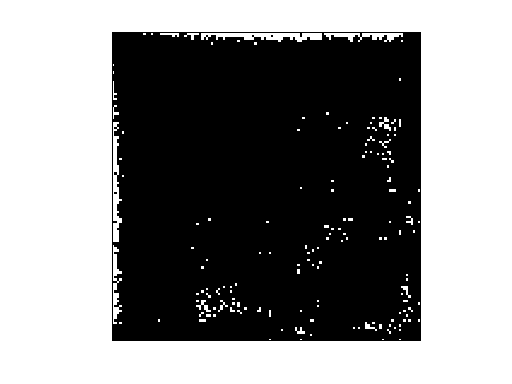}}	   &1,441,295	&3,099,940 &belgium \\
\hline
G3\_circuit&\mgape{\includegraphics[scale=0.025]{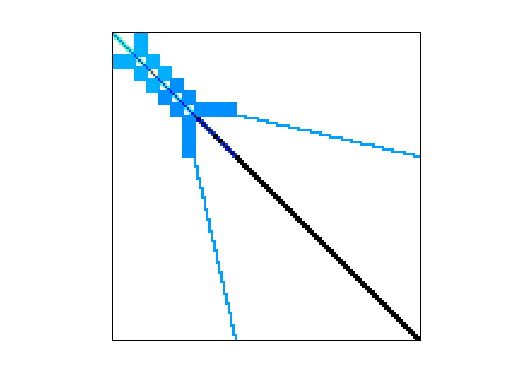}}  &1,585,478	   &4,623,152             &G3\\
\hline
roadNet-CA	&\mgape{\includegraphics[scale=0.025]{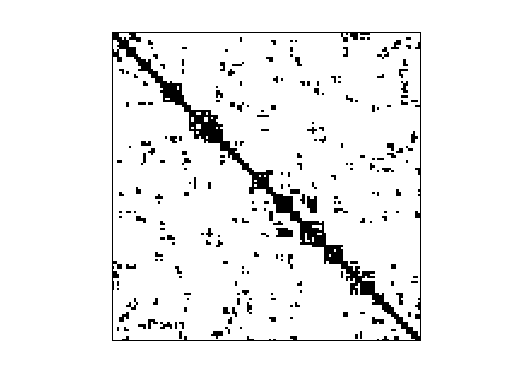}}       &1,971,281	&5,533,214 &roadNet\\
\hline
hugetrace-00020&\mgape{\includegraphics[scale=0.025]{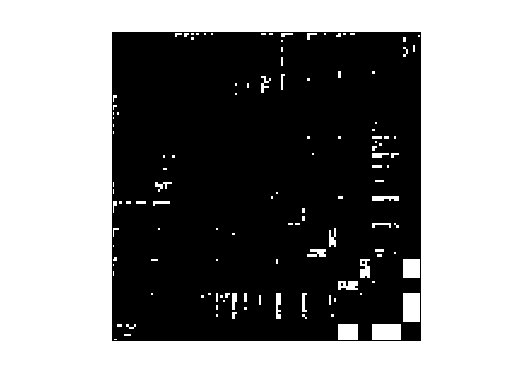}} &16,002,413&47,997,626  &hugetrace \\
\hline
hugetric-00020&\mgape{\includegraphics[scale=0.025]{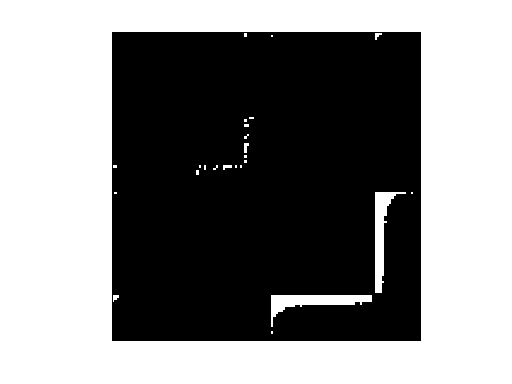}}&7,122,792 &21,361,554  &hugetric  \\
\hline
delaunay\_n24 &\mgape{\includegraphics[scale=0.025]{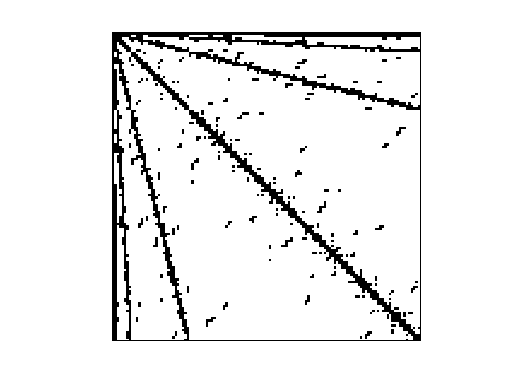}}     &16,777,216&100,663,202 &del24  \\
\hline
dielFilterV3real&\mgape{\includegraphics[scale=0.025]{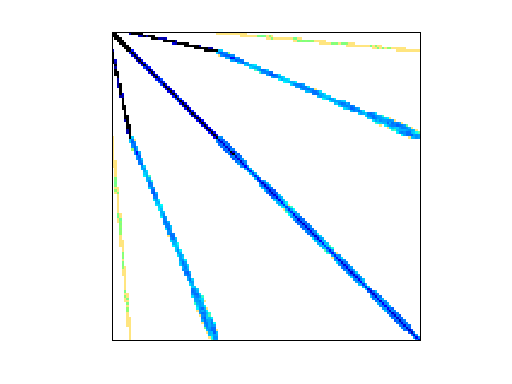}}   &1,102,824	&45,204,422 &diel\\
\hline
rgg\_n\_2\_24\_s0&\mgape{\includegraphics[scale=0.025]{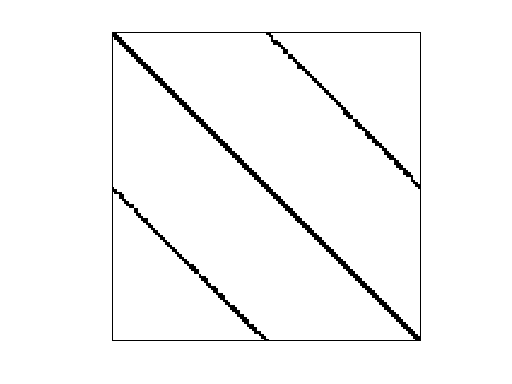}}  &16,777,216	&132,557,200 &rgg24\\
\hline
road\_usa&\mgape{\includegraphics[scale=0.025]{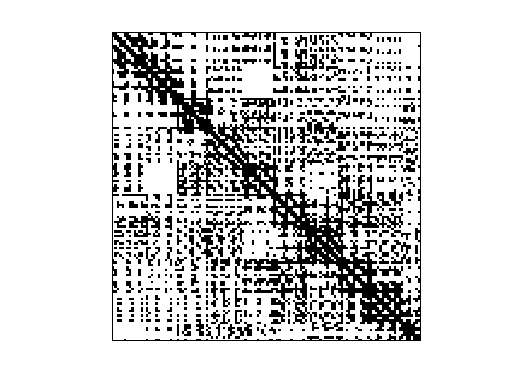}}	   &23,947,347	&57,708,624 &road\_usa\\
\hline
soc-LiveJournal1&\mgape{\includegraphics[scale=0.025]{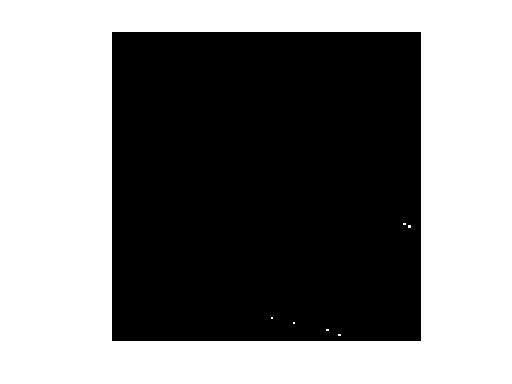}}  &4,847,571 &68,993,773  &soc       \\
\hline
ljournal-2008&\mgape{\includegraphics[scale=0.025]{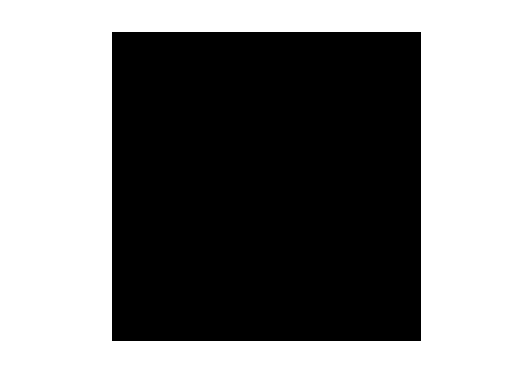}}  &5,363,260 &79,023,142  &ljournal  \\
\hline
kron\_g500-logn21&\mgape{\includegraphics[scale=0.025]{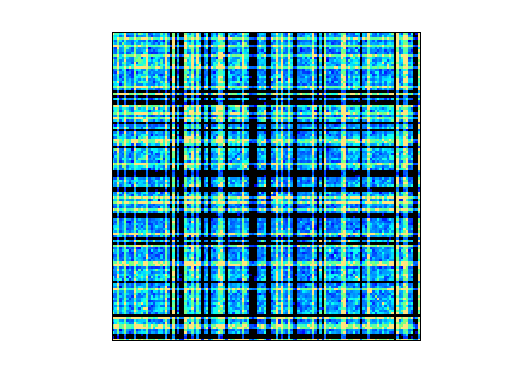}} &2,097,152	&182,082,942     &kron21\\
\hline
wikipedia-20070206&\mgape{\includegraphics[scale=0.025]{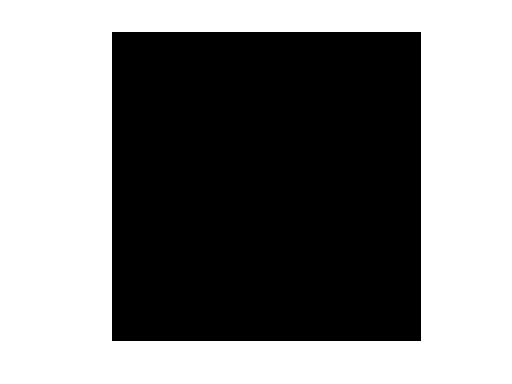}}&3,566,907 &45,030,389  &wikipedia \\
\hline
hollywood-2009&\mgape{\includegraphics[scale=0.025]{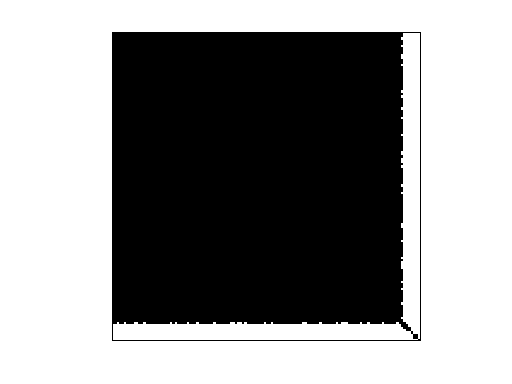}} &1,139,905	  &113,891,327  &hollywood	\\
\hline
flickr&\mgape{\includegraphics[scale=0.025]{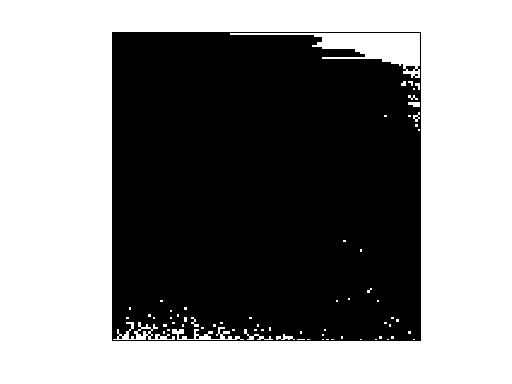}}	    &820,878	  &9,837,214     &flickr\\
\hline
soc-orkut&\mgape{\includegraphics[scale=0.025]{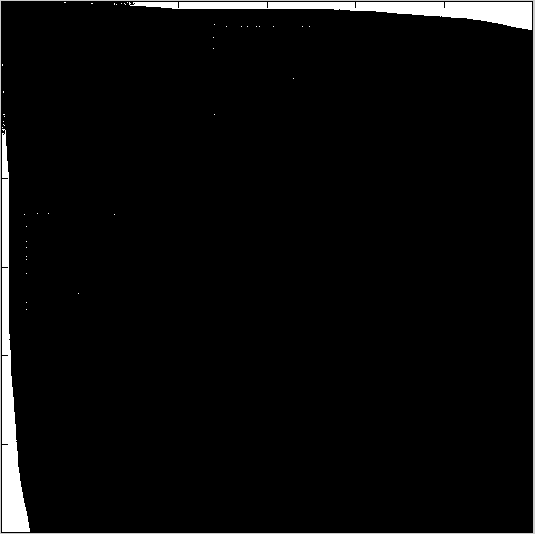}}	       &2,997,166	  &212,698,418   &soc-orkut\\
\hline
indochina-2004&\mgape{\includegraphics[scale=0.025]{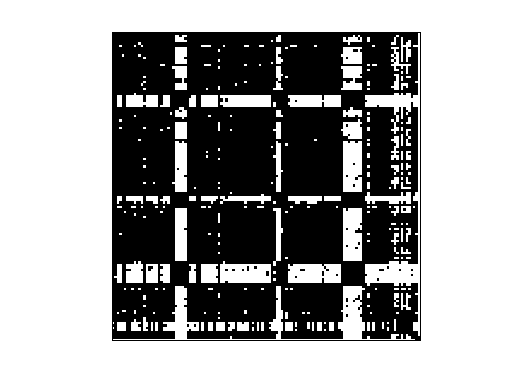}} &7,414,866 &194,109,311   &indochina\\
\hline
wb-edu&\mgape{\includegraphics[scale=0.025]{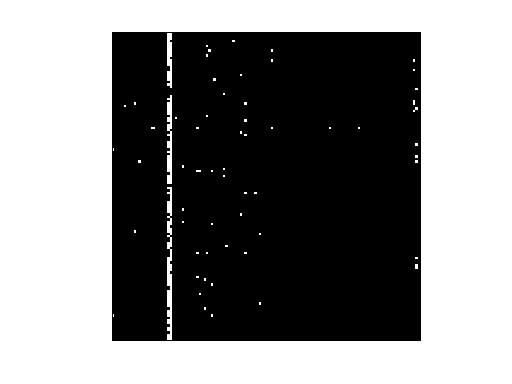}}	    &9,845,725	  &57,156,537    &wb-edu	\\
\hline
  \end{tabular}
\end{table}

In order to examine the performance of the proposed adaptive SpMV/SpMSpV framework,
we carry out a series of experiments on three typical NVIDIA GPUs with different micro-architectures,
including Tesla K40m,  P100 and  V100.
The Compute Unified Device Architecture (CUDA \cite{cuda10})  v10.1 is
employed to implement the SpMV/SpMSpV kernels and the proposed adaptive framework.
And GCC v4.8.5 and NVCC v10.1 with \texttt{-O3} flag are utilized to compile the code.
We use eighteen representative graph data sets for test,
as listed in Table \ref{table:dataset},
all of which have been widely utilized in many previous works  \cite{CombBLAS-2011,GraphMat-2015,GraphPad-2016,spmspv-sort-2015,spmspv-bucket-2017}.
Among them, the first nine can be used to represent high-diameter graphs,
while the last nine for low-diameter ones.
All these data sets are available from \cite{matrix-market} except \textit{soc-orkut},
which is from the Network Repository \cite{network}.
We use the single-precision data type in all tests and measure
the performance by averaging the results of ten repeated runs.

\begin{figure*}[!t]
\centering
\includegraphics[width=1.8\columnwidth]{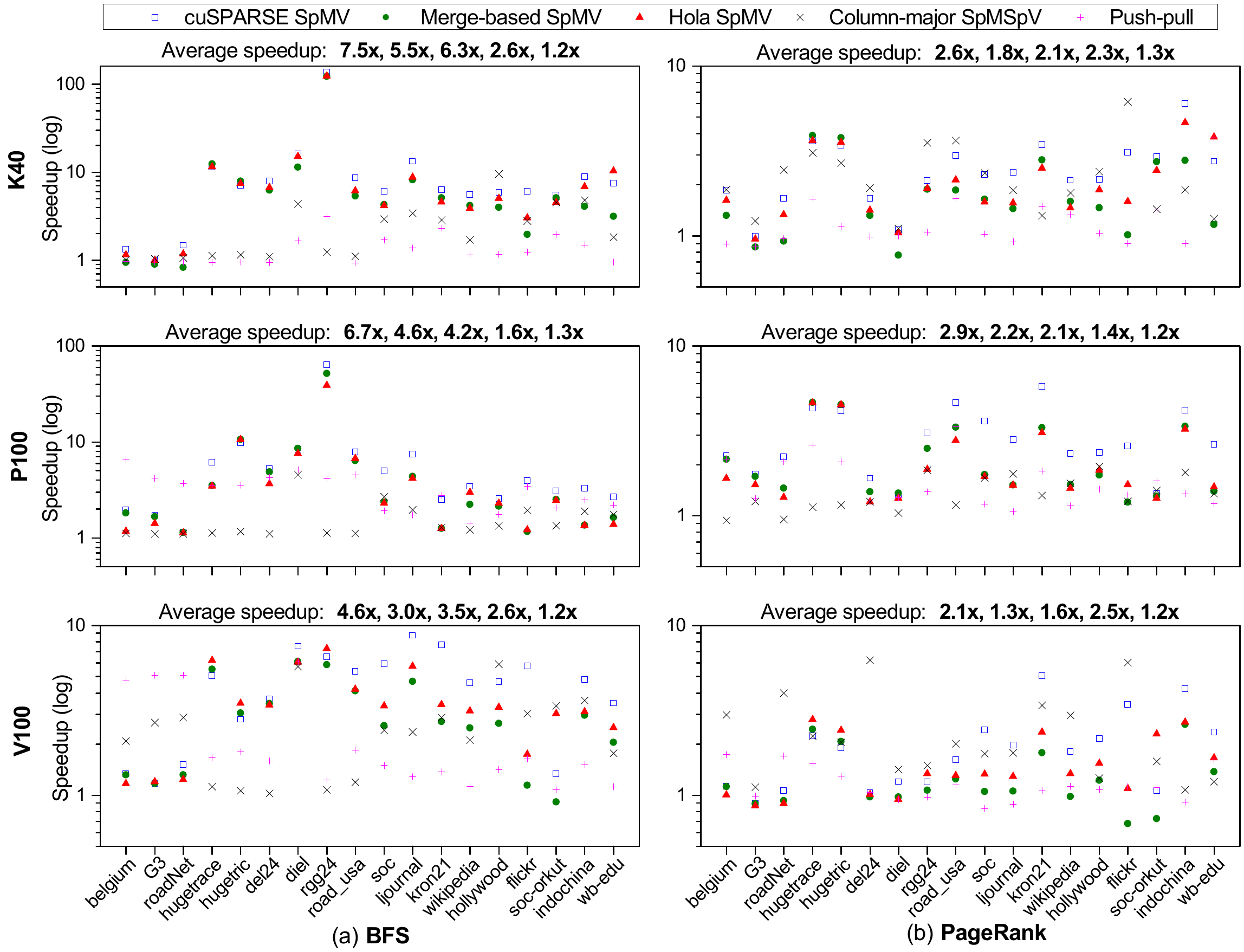}
\caption{Speedups of the adaptive framework over the reference SpMV/SpMSpV implementations in BFS and PageRank applications.}
\label{fig-application}
\end{figure*}

\subsection{Overall Performance}

In many real-world applications, the sparsity of input vector may dynamically change
during the program execution.
To investigate the flexibility and efficiency of the proposed adaptive framework in this case,
experiments are carried out by applying it to two real-world applications:
BFS and incremental PageRank \cite{pagerank}.
Instead of implementing a complex graph library from scratch,
we simulate the test by directly using the sparse vectors generated
from the existing graph library \cite{CombBLAS-2011, GraphMat-2015}.
Data sets of both high and low-diameter graphs in Table 1 are tested.
In the BFS tests, the first vertex is taken as the source, and the total number of iteration is
in the range of 514 to 2,803 for high-diameter graphs and 11 to 17 for low-diameter ones.
In the incremental PageRank tests, the total number of iteration is in the range of 27 to 288.

For comparison purpose, we also test the SpMV/SpMSpV implementations from
several previous state-of-the-art contributions
\cite{spmspv-push-pull-2018} \cite{spmspv-sort-2015} \cite{cusparse} \cite{Merge-spmv}
\cite{hola-spmv}.
Among them, the push-pull library \cite{spmspv-push-pull-2018}
uses an adaptive approach to switch between an SpMV kernel and an SpMSpV kernel.
And reference \cite{spmspv-sort-2015} relies on a single SpMSpV of the column-major type.
The other three works are all based on SpMV, taken from the cuSPARSE library \cite{cusparse},
the merge-based implementation \cite{Merge-spmv},
and the Hola SpMV \cite{hola-spmv}, respectively.
To make fare comparison,
the runtime overhead of the kernel selection process is also included in the test results.
The test results are summarize in Figure \ref{fig-application}.

From the figure, we can see that the proposed adaptive framework can outperform
all reference works across all three GPU platforms, in both BFS and PageRank applications.
As compared with the fixed SpMV kernels of cuSPARSE, merge-based SpMV, and  Hola SpMV,
the arithmetic average performance improvement is around 3.0x to 7.5x for BFS
and around 1.6x to 2.9x for PageRank, respectively.
With respect to the column-major SpMV, the sustained speedup of the adaptive framework
is around 1.4x to 2.6x for both applications.
We remark here that for the BFS tests with high-diameter graphs, the input vectors
are usually very sparse and the optimal kernels are usually column-major SpMV.
Our adaptive framework can still achieve a relatively large speedup in this case,
thanks to performance improvement brought {by
the adaptive selection} of different workload distribution and write-back strategies.
As compared to the adaptive SpMV/SpMSpV from the push-pull library, the proposed
adaptive framework performs very stable on all three GPU platforms in both applications.
The averaged performance gain is around 1.2x to 1.3x, which clearly demonstrates the advantage
of the adaptive framework.

\subsection{Prediction Accuracy}

\begin{figure}[!htb]
\centering
\includegraphics[width=0.85\columnwidth]{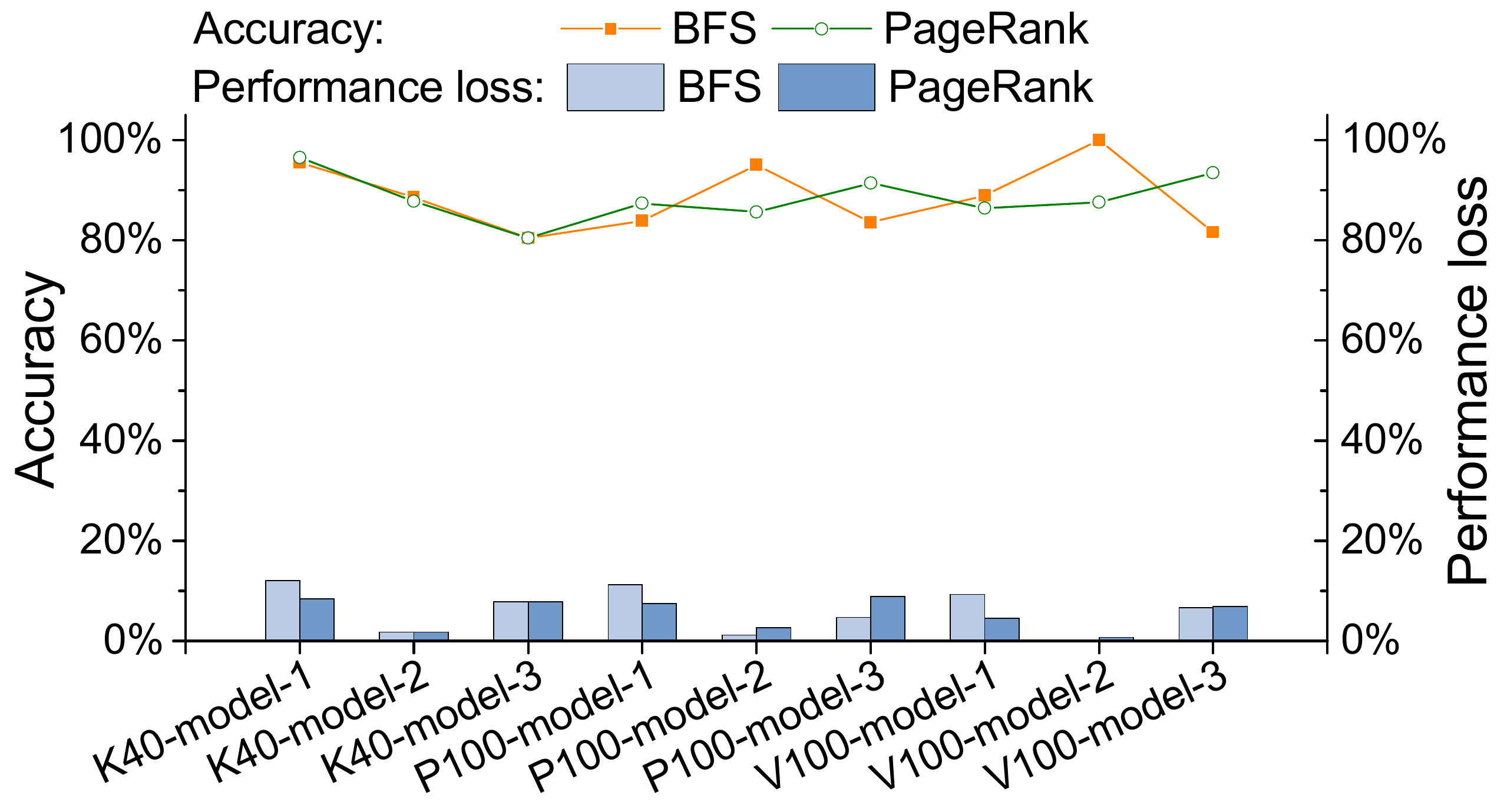}
\caption{The prediction accuracy and the performance loss due to misprediction of the three decision tree models on the three GPU platforms with the input vectors generated in BFS and {PageRank}.}
\label{fig-accuracy}
\end{figure}

To examine the prediction accuracy of the decision tree models in the BFS and incremental PageRank applications,
we further compare the performance of adaptive framework and the performance
of the fastest kernels selected through exhaustive search.
The test results are shown in Figure \ref{fig-accuracy},
from which we can observe that the prediction accuracy of the adaptive framework
can be maintained to a relatively high level of around 89.7\%, 90.7\% and 85.1\%
for the three decision tree models, respectively.
As a result, the performance loss due to misprediction is kept to a low level of
around 8.8\%, 1.3\% and 7.1\%  for the three decision tree models, respectively.

It is also of interest to note that in many cases the proposed adaptive framework can
achieve better performance than the reference implementations even when the models
make incorrect predictions.
The main reason is that the performance gap of different kernels is minor around the decision boundary,
and the wrong predictions for these cases still can lead to competitive performance.
The test results have suggested that the proposed adaptive framework can sustain satisfactory
performance on all the tested GPU platforms for various inputs and application scenarios.

\subsection{Prediction Overhead}

\begin{figure}[!htb]
\centering
\includegraphics[width=0.85\columnwidth]{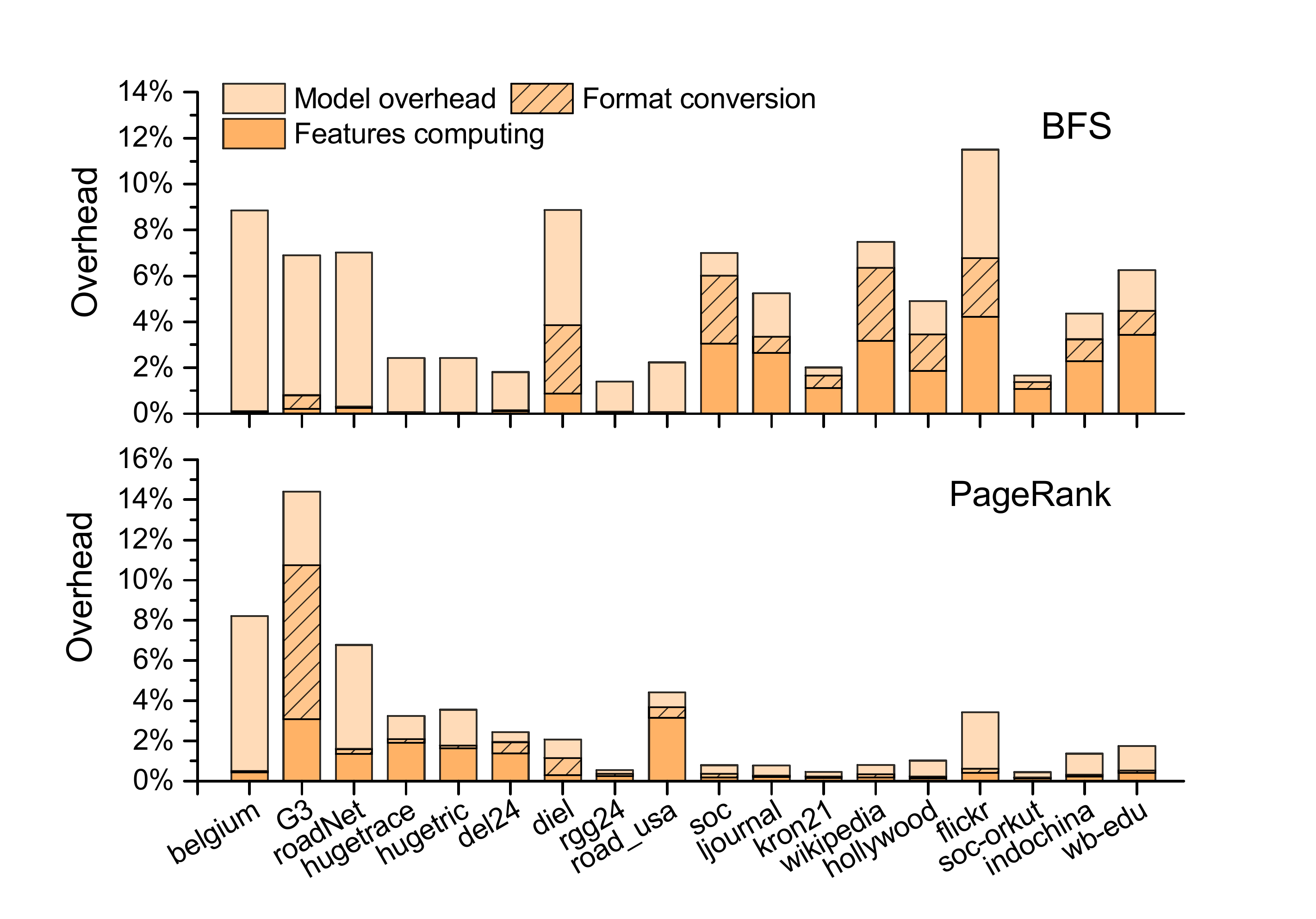}
\caption{The breakdown of the runtime prediction overhead of the adaptive framework on V100  with the input vectors generated in BFS and {PageRank}.}
\label{fig-overhead}
\end{figure}

In addition to the prediction accuracy, the overhead to make the prediction is also of practical importance.
We again use the BFS and incremental PageRank applications to do the analysis.
Since the results on different GPUs are similar, we take the results on V100 for demonstration purpose.
As discussed earlier, the overhead of the runtime prediction comes from
the computation of features, the conversion of vector format
and the evaluation of the decision tree rules.
A breakdown of the measured runtime overhead of each part is presented in Figure \ref{fig-overhead}.

For BFS, the overall overhead is less than 12\% and is averagely 7.2\% for all the data sets.
And the costs for feature computation, vector format conversion,
and model evaluation account for averagely 1.7\%, 1.1\% and 4.4\% of the overall runtime, respectively.
As for the incremental PageRank application, the overall overhead is less than 14\%
and is averagely 5.2\% for all the data sets.
And the overheads of three parts are on average 1.2\%, 0.7\% and 3.3\%, respectively.
The results also show that there do exist some cases with relative high overhead, such as
\textit{belgium}, \textit{G3} and \textit{flickr}. These situations usually occur {due
to the} relatively small size of the matrix and the fast execution of the selected kernel.
Despite these overheads, the proposed adaptive framework can still bring substantial performance
improvements, as shown in the previous test results.

\section{Discussion}
In this paper, we explore eight candidate kernels from three key factors based on the popular sparse matrix storage format CSR/CSC.
With the development of applications and hardware architectures,
more kernels or more factors may be introduced in the future.
If a new kernel implementation is devised based on, for example,
a new workload distribution method,
the adaptive framework is still applicable after necessary retraining.
The main effort is to incorporate the kernel into the framework and retrain the kernel selection models.
If a new factor is introduced, a new model can be added to select the alternatives under this factor.
When considering new sparse matrix storage formats,
the matrix storage format needs to be determined at the beginning of the iterations
to avoid the high overhead of runtime matrix format conversion.
Therefore, a separate matrix format selection model, such as SMAT \cite{SMAT-spmv},
can be added before the models of SpMV/SpMSpV selection.
Once the best matrix format is selected, the SpMV kernels of this format can be identified and used
as candidate kernels in the subsequent SpMV/SpMSpV selector.

\section{Related Work}
\label{related}

There are a large variety of studies on enhancing the performance of traditional SpMV.
We can categorize them into three classes:
(1) improving the performance of the classic CSR-based SpMV
\cite{Bell-2008-spmv, Ashari-sc14-spmv,Greathouse-sc14-spmv,cusparse,Merge-spmv,naive-spmv,hola-spmv},
(2) designing new sparse matrix formats for more efficient SpMV computations
\cite{CSR5, new-format-CSX, new-format-clspmv,ELL-liu-2013, slice-ELL-2010, 2018-CVR-spmv},
and (3) selecting the optimal format
according to the characteristic of the sparse matrices
\cite{SMAT-spmv, 2016-icpp-spmv-select, adaptive-1, dnn-spmv, spmv-adaptive-1, spmv-adaptive-2}.
{The adaptive framework proposed in this paper takes full advantages of the CSR-based SpMV on GPUs.}
Early works includes CSR-scalar \cite{Bell-2008-spmv}, CSR-adaptive
\cite{Ashari-sc14-spmv}, and CSR-bin \cite{Greathouse-sc14-spmv}. Recently,
merge-based SpMV \cite{Merge-spmv} and Hola-SpMV
\cite{hola-spmv} were proposed to further deal with issues such as irregular memory access and
load imbalanced work distribution.
{In this paper,} both the direct and the load-balanced work-distribution SpMV kernels
are explored and incorporated in the proposed framework, which can work well
for matrices with both uniform and non-uniform nonzero distributions.
Besides, compared with the optimal sparse matrix format selection works for SpMV
\cite{SMAT-spmv,2016-icpp-spmv-select,adaptive-1,dnn-spmv,spmv-adaptive-1,spmv-adaptive-2},
where the input vector is always dense, we focus more on SpMV/SpMSpV kernel selection when the sparsity
of the input vector varies during the execution.

Early driven forces on the performance optimization of SpMSpV were mainly from graph analytics.
For example, CombBLAS \cite{CombBLAS-2011} is a graph computation library
that provides a number of graph algorithms.
Many of these algorithms rely on a vector-driven column-major SpMSpV kernel and can
work well on both serial and parallel CPU platforms.
The multi-threaded SpMSpV in CombBLAS was later enhanced with a novel bucket algorithm
to provide more efficient writeback of partial sums \cite{spmspv-bucket-2017}.
GraphMat \cite{GraphMat-2015} is another graph computation library built with
a matrix-driven column-major SpMSpV algorithm for single-CPU environments.
It is worth noting that the matrix-driven approach is usually more suitable for row-major SpMSpV on GPUs,
instead of the column-major one in GraphMat,  as was analysed in our paper.
GraphPad \cite{GraphPad-2016} is an extension of GraphMat to support multi-node CPU computing.
On GPUs, a vector-driven column-major SpMSpV algorithm was proposed in reference \cite{spmspv-sort-2015},
which utilized a merge-based method to keep the workload balanced distributed
and a sort-based writing-back method to reduce the writing conflicts.
In this paper, both matrix-driven row-major SpMSpV and vector-driven column-major SpMSpV
are explored as the candidate solutions of the SpMSpV computation on GPUs.
Besides, we also expand the candidate kernels from the workload distribution and writing-back perspectives.
In total, six  SpMSpV kernels are explored and integrated in the proposed
adaptive framework to help adapt to inputs with different characteristics.

Adaptive kernel selection has been a promising direction to be explored over the years\cite{adaptive-2, adaptive-3, adaptive-4}.
A large variety of studies have been done for adaptive SpMV
\cite{SMAT-spmv, 2016-icpp-spmv-select, spmv-model-likenli, adaptive-1, dnn-spmv, spmv-adaptive-1, spmv-adaptive-2}
and adaptive spGEMM \cite{sw-spMM, iaspgemm}.
All of these works focus on the adaptive selection of sparse matrix formats.
In this paper, we mainly focus on the adaptive selection of SpMV/SpMSpV kernels for input vectors with both fixed
and dynamically varied sparsity. To illustrate the idea, only a single sparse matrix format
can be utilized as a demonstration.
Our work is in fact orthogonal to the adaptive selection of different matrix storage formats.
A most similar work to ours is the adaptive framework from the push-pull library \cite{spmspv-push-pull-2018},
which can automatically select the vector-driven column-major SpMSpV and the load-balanced SpMV kernel
by using a simple heuristic method.
In the heuristic method, {the kernel switch point is set to a fixed value of the vector sparsity by user},
which is indeed not an accurate prediction and is not flexible with the change of hardware.
Compared with \cite{spmspv-push-pull-2018}, the proposed adaptive framework can explore
a larger SpMV/SpMSpV kernel space consisting of eight kernels by considering
several performance related factors.
With the help of machine learning models, it can automatically select
the optimal SpMV/SpMSpV kernels with low overhead and high accuracy,
and can easily generalize to other hardware platforms.

\section{Conclusion and Future Work}
\label{conclusion}

In this paper, an adaptive SpMV/SpMSpV framework was proposed to automatically select the appropriate
kernel based on a low-overhead machine learning model for the input vector with different sparsity.
Based on systematic analysis on key factors such as the computing pattern, workload distribution method and
write-back strategy, eight candidate SpMV/SpMSpV kernels were encapsulated into the framework to achieve high performance
in a seamless manner. A decision tree based kernel selector
was designed to choose the kernel and adapt with the varieties of both the input and hardware.
Experiments show that the proposed adaptive SpMV/SpMSpV framework can work well on three typical
GPU platforms and substantially outperform the previous state-of-the-art in real applications.
As an attempt to enable adaptive high-performance SpMV/SpMSpV on modern GPU platform,
this work shows promising results and great potentials.
In the future, we plan to further explore and include more feasible computation patterns
and implementation schemes into the framework and make it more portable on a broader range
of hardware platforms.

\section*{Acknowledgments}

The work was supported in part by National Key R\&D Plan of China (Grant No. 2016YFB0200603), 
Guangdong Key-Area R\&D Program (\#2019B121204008), Beijing Natural Science
Foundation (Grant No. JQ18001) and Beijing Academy of Artificial Intelligence.

\bibliographystyle{IEEEtran}
\bibliography{SMSV}

% Generated by IEEEtran.bst, version: 1.14 (2015/08/26)
\begin{thebibliography}{10}
\providecommand{\url}[1]{#1}
\csname url@samestyle\endcsname
\providecommand{\newblock}{\relax}
\providecommand{\bibinfo}[2]{#2}
\providecommand{\BIBentrySTDinterwordspacing}{\spaceskip=0pt\relax}
\providecommand{\BIBentryALTinterwordstretchfactor}{4}
\providecommand{\BIBentryALTinterwordspacing}{\spaceskip=\fontdimen2\font plus
\BIBentryALTinterwordstretchfactor\fontdimen3\font minus
  \fontdimen4\font\relax}
\providecommand{\BIBforeignlanguage}[2]{{%
\expandafter\ifx\csname l@#1\endcsname\relax
\typeout{** WARNING: IEEEtran.bst: No hyphenation pattern has been}%
\typeout{** loaded for the language `#1'. Using the pattern for}%
\typeout{** the default language instead.}%
\else
\language=\csname l@#1\endcsname
\fi
#2}}
\providecommand{\BIBdecl}{\relax}
\BIBdecl

\bibitem{saad-book}
Y.~Saad, \emph{Iterative Methods for Sparse Linear Systems}.\hskip 1em plus
  0.5em minus 0.4em\relax {SIAM}, 2003.

\bibitem{bfs}
A.~Bulu{\c{c}} and K.~Madduri, ``Parallel breadth-first search on distributed
  memory systems,'' in \emph{Proceedings of the Conference on High Performance
  Computing Networking, Storage and Analysis}, 2011, pp. 65:1--65:12.

\bibitem{pagerank}
S.~Ramachandran, ``Incremental {P}age{R}ank acceleration using {Sparse}
  {Matrix}-{Sparse} {Vector} {Multiplication},'' Ph.D. dissertation, The Ohio
  State University, 2016.

\bibitem{GraphMat-2015}
N.~Sundaram, N.~Satish, M.~M.~A. Patwary, S.~R. Dulloor, M.~J. Anderson, S.~G.
  Vadlamudi, D.~Das, and P.~Dubey, ``{GraphMat}: High performance graph
  analytics made productive,'' \emph{Proceedings of the VLDB Endowment},
  vol.~8, no.~11, pp. 1214--1225, 2015.

\bibitem{cpu-spmv-1999-sc}
A.~Pinar and M.~T. Heath, ``Improving performance of sparse matrix-vector
  multiplication,'' in \emph{Proceedings of the 1999 ACM/IEEE Conference on
  Supercomputing}, 1999.

\bibitem{cpu-spmv-2009}
A.~Bulu{\c{c}}, J.~T. Fineman, M.~Frigo, J.~R. Gilbert, and C.~E. Leiserson,
  ``Parallel sparse matrix-vector and matrix-transpose-vector multiplication
  using compressed sparse blocks,'' in \emph{Proceedings of Symposium on
  Parallelism in Algorithms and Architectures}, 2009, pp. 233--244.

\bibitem{cpu-spmv-2010-cgo}
K.~Kourtis, G.~Goumas, and N.~Koziris, ``Exploiting compression opportunities
  to improve {SpMxV} performance on shared memory systems,'' \emph{ACM
  Transactions on Architecture and Code Optimization}, vol.~7, no.~3, pp.
  16:1--16:31, 2010.

\bibitem{cpu-spmv-2011-ipdps}
A.~Bulu{\c{c}}, S.~Williams, L.~Oliker, and J.~Demmel, ``Reduced-bandwidth
  multithreaded algorithms for sparse matrix-vector multiplication,'' in
  \emph{Proceedings of International Parallel and Distributed Processing
  Symposium}, 2011, pp. 721--733.

\bibitem{pOSKI-2012}
J.-H. Byun, R.~Lin, J.~W. Demmel, and K.~A. Yelick, ``{pOSKI}: Parallel
  optimized sparse kernel interface library,'' in \emph{Technical Report},
  2012.

\bibitem{2018-CVR-spmv}
B.~Xie, J.~Zhan, X.~Liu, W.~Gao, Z.~Jia, X.~He, and L.~Zhang, ``{CVR}:
  Efficient vectorization of {SpMV} on {X}86 processors,'' in \emph{Proceedings
  of International Symposium on Code Generation and Optimization}, 2018, pp.
  149--162.

\bibitem{Bell-2008-spmv}
N.~Bell and M.~Garland, ``Implementing sparse matrix-vector multiplication on
  throughput-oriented processors,'' in \emph{Proceedings of the Conference on
  High Performance Computing Networking, Storage and Analysis}, 2009, pp.
  18:1--18:11.

\bibitem{Ashari-sc14-spmv}
A.~Ashari, N.~Sedaghati, J.~Eisenlohr, S.~Parthasarathy, and P.~Sadayappan,
  ``Fast sparse matrix-vector multiplication on {GPUs} for graph
  applications,'' in \emph{Proceedings of the Conference on High Performance
  Computing Networking, Storage and Analysis}, 2014, pp. 781--792.

\bibitem{Greathouse-sc14-spmv}
J.~L. Greathouse and M.~Daga, ``Efficient sparse matrix-vector multiplication
  on {GPUs} using the {CSR} storage format,'' in \emph{Proceedings of the
  Conference on High Performance Computing Networking, Storage and Analysis},
  2014, pp. 769--780.

\bibitem{CSR5}
W.~Liu and B.~Vinter, ``{CSR5:} an efficient storage format for cross-platform
  sparse matrix-vector multiplication,'' in \emph{Proceedings of International
  Conference on Supercomputing}, 2015, pp. 339--350.

\bibitem{Merge-spmv}
D.~Merrill and M.~Garland, ``Merge-based sparse matrix-vector multiplication
  ({SpMV}) using the {CSR} storage format,'' in \emph{ACM SIGPLAN Symposium on
  Principles and Practice of Parallel Programming}, 2016, pp. 43:1--43:2.

\bibitem{naive-spmv}
M.~Steinberger, A.~Derlery, R.~Zayer, and H.~P. Seidel, ``How naive is naive
  {SpMV} on the {GPU}?'' in \emph{Proceedings of High Performance Extreme
  Computing Conference}, 2016, pp. 1--8.

\bibitem{hola-spmv}
M.~Steinberger, R.~Zayer, and H.~P. Seidel, ``Globally homogeneous, locally
  adaptive sparse matrix-vector multiplication on the {GPU},'' in
  \emph{Proceedings of International Conference on Supercomputing}, 2017, pp.
  13:1--13:11.

\bibitem{new-format-CSX}
K.~Kourtis, V.~Karakasis, G.~Goumas, and N.~Koziris, ``{CSX}: An extended
  compression format for {SpMV} on shared memory systems,'' \emph{ACM SIGPLAN
  Notices}, vol.~46, no.~8, pp. 247--256, 2011.

\bibitem{new-format-clspmv}
B.~Y. Su and K.~Keutzer, ``{clSpMV}:a cross-platform {OpenCL SpMV} framework on
  {GPUs},'' in \emph{Proceedings of International Conference on
  Supercomputing}, 2012, pp. 353--364.

\bibitem{SMAT-spmv}
J.~Li, G.~Tan, M.~Chen, and N.~Sun, ``{SMAT}: An input adaptive auto-tuner for
  sparse matrix-vector multiplication,'' in \emph{Proceedings of {ACM SIGPLAN}
  Conference on Programming Language Design and Implementation}, 2013, pp.
  117--126.

\bibitem{dnn-spmv}
Y.~Zhao, J.~Li, C.~Liao, and X.~Shen, ``Bridging the gap between deep learning
  and sparse matrix format selection,'' vol.~53, no.~1, pp. 94--108, 2018.

\bibitem{libsvm}
C.-C. Chang and C.-J. Lin, ``{LIBSVM}: A library for support vector machines,''
  \emph{ACM Transactions on Intelligent Systems and Technology}, vol.~2, no.~3,
  p.~27, 2011.

\bibitem{CombBLAS-2011}
A.~Bulu{\c{c}} and J.~R. Gilbert, ``The combinatorial {BLAS}: Design,
  implementation, and applications,'' \emph{Proceedings of the International
  Journal of High Performance Computing Applications}, vol.~25, no.~4, pp.
  496--509, 2011.

\bibitem{GraphPad-2016}
M.~J. Anderson, N.~Sundaram, N.~Satish, M.~M.~A. Patwary, T.~L. Willke, and
  P.~Dubey, ``{GraphPad}: Optimized graph primitives for parallel and
  distributed platforms,'' in \emph{Proceedings of the International Parallel
  and Distributed Processing Symposium}, 2016, pp. 313--322.

\bibitem{spmspv-bucket-2017}
A.~Azad and A.~Bulu{\c{c}}, ``A work-efficient parallel sparse matrix-sparse
  vector multiplication algorithm,'' in \emph{Proceedings of the International
  Parallel and Distributed Processing Symposium}, 2017, pp. 688--697.

\bibitem{spmspv-sort-2015}
C.~Yang, Y.~Wang, and J.~D. Owens, ``Fast sparse matrix and sparse vector
  multiplication algorithm on the {GPU},'' in \emph{Proceedings of the
  International Parallel and Distributed Processing Symposium Workshop}, 2015,
  pp. 841--847.

\bibitem{p100-whitepaper}
{NVIDIA}, ``{NVIDIA Tesla GP100 Pacal whitepaper},'' 2016.

\bibitem{v100-whitepaper}
------, ``{NVIDIA Tesla V100 GPU Archtecture},'' 2017.

\bibitem{direction-bfs}
S.~Beamer, K.~Asanovi\'{c}, and D.~Patterson, ``Direction-optimizing
  breadth-first search,'' \emph{Scientific Programming}, vol.~21, no. 3-4, pp.
  137--148, 2013.

\bibitem{spmspv-push-pull-2018}
C.~Yang, A.~Bulu{\c{c}}, and J.~D. Owens, ``Implementing push-pull efficiently
  in {GraphBLAS},'' in \emph{Proceedings of the International Conference on
  Parallel Processing}, 2018, pp. 89:1--89:11.

\bibitem{gini}
J.~Preusse, ``Fairness on the web: Alternatives to the power law,'' in
  \emph{Proceedings of the 3rd Annual ACM Web Science Conference}, 2012, pp.
  175--184.

\bibitem{ada-trsv}
N.~Ahmad, B.~Yilmaz, and D.~Unat, ``A prediction framework for fast sparse
  triangular solves.'' in \emph{: Malawski M., Rzadca K. (eds) Euro-Par 2020:
  Parallel Processing. Euro-Par 2020. Lecture Notes in Computer Science}, vol.
  12247, 2020.

\bibitem{matrix-market}
T.~A. Davis and Y.~Hu, ``The {U}niversity of {F}lorida {S}parse {M}atrix
  {C}ollection,'' \emph{ACM Transactions on Mathematical Software}, vol.~38,
  no.~1, pp. 1:1--1:25, 2011.

\bibitem{sklearn}
F.~Pedregosa, G.~Varoquaux, A.~Gramfort, V.~Michel, B.~Thirion, and et~al,
  ``Scikit-learn: Machine learning in {P}ython,'' \emph{Journal of Machine
  Learning Research}, vol.~12, pp. 2825--2830, 2011.

\bibitem{cuda10}
\BIBentryALTinterwordspacing
``The programming guide to the {CUDA} model and interface.'' 2019. [Online].
  Available:
  \url{https://docs.nvidia.com/cuda/archive/10.1/cuda-c-programming-guide/index.html}
\BIBentrySTDinterwordspacing

\bibitem{network}
R.~A. Rossi and N.~K. Ahmed, ``An interactive data repository with visual
  analytics,'' \emph{SIGKDD Explor. Newsl.}, vol.~17, no.~2, pp. 37--41, 2016.

\bibitem{cusparse}
\BIBentryALTinterwordspacing
``{The {API} reference guide for {cuSPARSE}, the {CUDA} sparse matrix
  library}.'' 2019. [Online]. Available:
  \url{https://docs.nvidia.com/cuda/archive/10.1/cusparse/index.html}
\BIBentrySTDinterwordspacing

\bibitem{ELL-liu-2013}
X.~Liu, M.~Smelyanskiy, E.~Chow, and P.~Dubey, ``Efficient sparse matrix-vector
  multiplication on {X86}-based many-core processors,'' in \emph{Proceedings of
  International Conference on Supercomputing}, 2013, pp. 273--282.

\bibitem{slice-ELL-2010}
A.~Monakov, A.~Lokhmotov, and A.~Avetisyan, ``Automatically tuning sparse
  matrix-vector multiplication for {GPU} architectures,'' in \emph{Proceedings
  of International Conference on High-Performance Embedded Architectures and
  Compilers}, 2010, pp. 111--125.

\bibitem{2016-icpp-spmv-select}
A.~Benatia, W.~Ji, Y.~Wang, and F.~Shi, ``Sparse matrix format selection with
  multiclass {SVM} for {SpMV} on {GPU},'' in \emph{Proceddings of International
  Conference on Parallel Processing}, 2016, pp. 496--505.

\bibitem{adaptive-1}
H.~Cui, S.~Hirasawa, H.~Takizawa, and H.~Kobayashi, ``A code selection
  mechanism using deep learning,'' in \emph{Proceedings of the International
  Symposium on Embedded Multicore/Many-core Systems-on-Chip}, 2016, pp.
  385--392.

\bibitem{spmv-adaptive-1}
Y.~{Zhao}, W.~{Zhou}, X.~{Shen}, and G.~{Yiu}, ``Overhead-conscious format
  selection for {SpMV}-based applications,'' in \emph{Proceedings of the
  International Parallel and Distributed Processing Symposium}, 2018, pp.
  950--959.

\bibitem{spmv-adaptive-2}
I.~{Nisa}, C.~{Siegel}, A.~S. {Rajam}, A.~{Vishnu}, and P.~{Sadayappan},
  ``Effective machine learning based format selection and performance modeling
  for {SpMV} on {GPUs},'' in \emph{Proceedings of the International Parallel
  and Distributed Processing Symposium Workshops}, 2018, pp. 1056--1065.

\bibitem{adaptive-2}
Y.~Ding, J.~Ansel, K.~Veeramachaneni, X.~Shen, U.-M. O'Reilly, and
  S.~Amarasinghe, ``Autotuning algorithmic choice for input sensitivity,''
  \emph{ACM SIGPLAN Notices}, vol.~50, 2014.

\bibitem{adaptive-3}
P.~Tillet and D.~Cox, ``Input-aware auto-tuning of compute-bound {HPC}
  kernels,'' in \emph{Proceedings of the International Conference for High
  Performance Computing, Networking, Storage and Analysis}, 2017, pp.
  43:1--43:12.

\bibitem{adaptive-4}
S.~Muralidharan, M.~Shantharam, M.~Hall, M.~Garland, and B.~Catanzaro,
  ``{Nitro}: A framework for adaptive code variant tuning,'' in
  \emph{Proceedings of the International Parallel and Distributed Processing
  Symposium}, 2014, pp. 501--512.

\bibitem{spmv-model-likenli}
K.~Li, W.~Yang, and K.~Li, ``Performance analysis and optimization for {SpMV}
  on {GPU} using probabilistic modeling,'' \emph{IEEE Transactions on Parallel
  and Distributed Systems}, vol.~26, no.~1, pp. 196--205, 2014.

\bibitem{sw-spMM}
Y.~Chen, K.~Li, W.~Yang, G.~Xiao, X.~Xie, and T.~Li, ``Performance-aware model
  for sparse matrix-matrix multiplication on the {Sunway TaihuLight}
  supercomputer,'' \emph{IEEE Transactions on Parallel and Distributed
  Systems}, vol.~30, no.~4, pp. 923--938, 2018.

\bibitem{iaspgemm}
Z.~Xie, G.~Tan, W.~Liu, and N.~Sun, ``{IA-SpGEMM}: An input-aware auto-tuning
  framework for parallel sparse matrix-matrix multiplication,'' in
  \emph{Proceedings of the ACM International Conference on Supercomputing},
  2019, pp. 94--105.

\end{thebibliography}

\end{document}